\documentstyle[epsf,12pt,axodraw]{article}
\newcommand{\z}{&\hspace*{-8pt}}
\newcommand{\Li}{{\rm Li}}

\newcommand{\trg}[3]{
\begin{picture}(35,30)(5,13)
\Line(5,5)(20,30)
\Line(35,5)(20,30)
\Line(5,5)(35,5)
\Line(5,5)(3,3)
\Line(35,5)(38,3)
\Line(20,30)(20,32)
\Text(20,0)[]{$\scriptstyle #1$}
\Text(13,25)[r]{$\scriptstyle #2$}
\Text(27,25)[l]{$\scriptstyle #3$}
\end{picture}
                    }

\newcommand{\trgleft}[3]{
\begin{picture}(35,30)(5,13)
\Line(5,5)(20,30)
\Line(35,5)(20,30)
\Line(5,5)(35,5)
\DashLine(5,5)(-10,5){2}
\Line(-10,5)(-13,3)
\Line(35,5)(38,3)
\Line(20,30)(20,32)
\Text(-3,0)[]{$\scriptstyle -1$}
\Text(20,0)[]{$\scriptstyle #1$}
\Text(13,25)[r]{$\scriptstyle #2$}
\Text(27,25)[l]{$\scriptstyle #3$}
\end{picture}
                        }

\begin{document}

\begin{flushright}
BI-TP-98/20 \\
\end{flushright}

\begin{center}

\vspace*{40mm}

{\Large \bf Analytic two-loop results for selfenergy- and vertex-type diagrams
            with one non-zero mass}

\vskip 10mm

J.~Fleischer
\footnote{~E-mail: fleischer@physik.uni-bielefeld.de}
A.~V.~Kotikov
\footnote{~Particle Physics Lab.,
Joint Institute for Nuclear Research, 141980, Dubna (Moscow Region), Russia}
\footnote{~E-mail: kotikov@sunse.jinr.dubna.su}
\footnote{~Supported by Volkswagen-Stiftung under I/71~293 
and RFFI uncer 98-02-16923}
and
O.~L.~Veretin
\footnote{~E-mail:veretin@physik.uni-bielefeld.de}
\footnote{~Supported by BMBF under 05~7BI92P(9)}

\vskip 10mm

{\it ~Fakult\"at f\"ur Physik, Universit\"at Bielefeld,
D-33615 Bielefeld, Germany.}

\begin{abstract}

For a large class of two-loop selfenergy- and vertex-type diagrams 
with only one non-zero mass ($M$) and the vertices also with only one 
non-zero external momentum squared ($q^2$) the first few expansion 
coefficients are calculated
by the large mass expansion. This allows to `guess' the general structure
of these coefficients and to
verify them in terms of certain classes of `basis elements', which are
essentially harmonic sums. 
Since for this case with only one non-zero mass the large mass expansion and
the Taylor series in terms of $q^2$ are identical, this approach yields
analytic expressions of the Taylor coefficients, from which the diagram can
be easily evaluated numerically in a large domain of the complex $q^2-$plane 
by well known methods. 
It is also possible to sum the Taylor series and present the results
in terms of polylogarithms.

\end{abstract}
\end{center}

\\

\vfill

\thispagestyle{empty}
\setcounter{page}0
\newpage

\section{Introduction}

  Higher loop calculations present great difficulties
especially in the presence of masses.
When there is a large number of parameters 
(masses and external momenta) in the problem,
analytic expressions can hardly be obtained
beyond one-loop. In such cases one has to resort
to numerical or approximative methods.
In particular the rather large spectrum of masses in the 
Standard Model requires numerous and different
approximation schemes (e.g. small fermion masses $m_f^2/m_Z^2\ll 1$,
close vector boson masses $(m_Z^2-m_W^2)/m_Z^2\ll 1$,
large top mass $m_Z^2/m_t^2\ll 1$ etc.) and
often a problem is reduced to diagrams with one mass.

  In this work we consider the evaluation of 
2-loop diagrams of self-energy and vertex type with one
non-zero mass and all others zero.

  For the analytic evaluation of `master integrals'
with masses various techniques have been developed. 
Such calculations in QED are known
since \cite{Kaellen}.
Using dispersion relations 2-loop 2-point 
integrals relevant for QED and QCD were obtained
in \cite{BroadhurstZP47}. Results for a class of
2-loop self energies are given in \cite{SE2}-\cite{Ghinculov}.
The method of differential equations was developed
in \cite{DEM}.
Some recent results, applying this method 
for 2-loop vertex functions,
are presented in \cite{FKV}. At the 3-loop order
analytic results are known only for vacuum
bubbles so far \cite{bubbles}.

  Very powerful approaches for the evaluation of
diagrams, which are in fact our starting point,
are the asymptotic expansion \cite{asymptotic}
and/or Taylor expansion \cite{ft,DT} in an external momentum squared, 
which are identical if only one non-zero mass is involved.
These methods allow to expand a diagram
in a series with coefficients, which in many cases
can be represented in a relatively simple analytic form.
Analytic continuation by means of a mapping in the plane
of momentum squared allows to calculate the diagrams on 
their cut. For propagator type diagrams such expansion
was successfully applied in \cite{propagator,propagatorbis} while
for vertex functions it is discussed in \cite{ft}.

  So far the method of expansion is considered 
as semianalytic in the sence that only a limited
number of coefficients can be obtained explicitely.
In this paper, however, we want to go one step further.
We calculate the first few coefficients of the
expansion of a diagram with only one non-zero mass
by means of the asymptotic expansion \cite{asymptotic}
method. The method of differential equations \cite{DEM}
then yields an idea, like in \cite{FKV}, what the general 
analytic form of these coefficients might be, providing
some `basis' in terms of which they might be expressed.
The Ansatz of equating the explicit coefficients obtained 
from the asymptotic expansion to a linear combination
in terms of the basis elements, yields a system of linear
equations, which can be solved to yield the desired representation
of the coefficients. 

   The main problem in this approach is the choice of the 
basis elements. We start from so-called harmonic sums
which are particularly relevant for moments of
structure functions in QCD (e.g. \cite{Yndurain}).
These functions are directly related to
(generalized) polylogarithms \cite{polylogarithms}
and therefore it is not surprising that they appear
in the analysis of massive diagrams (see also recent research 
about these sums \cite{VermRh}). 
However, not every massive
diagram even of self-energy type can be expressed
in terms of these sums. In particular, if a diagram 
possesses a cut with three massive particles in the intermediate
state then it is rather a subject of elliptic
integrals than polylogarithmic ones \cite{BroadhurstZP47,Bauberger}.
The method should work in this case as well
provided one finds the relevant class of basis elements.
However, in this paper we do not consider diagrams
of such type and concentrate ourselves on those
with one- and/or two-massive-particle ($m$- and $2m$-) cuts.
We give the results for all selfenergies and a certain number
of vertices with $m$- and $2m$-cuts. In particular
we include diagrams relevant for $Z$-decay.

  The paper is organized as follows: after introducing some
notation in Sect. 2, we explain in Sect. 3 the main idea of
our approach in terms of some simple examples. Sect. 4 contains
a `straightforward' prescription for the generalization of our
basis elements, while in Sect.5 we demonstrate how to find more
complicated ones occurring in $2m$- cuts. The differential equation
method (DEM) in Sect. 6 demonstrates a further method to explore the 
form of basis elements. Sect. 7 contains a proposal of how to 
perform the large $q^2$-expansion by means of the Sommerfeld-Watson
transformation, providing an explicit example. The Appendices contain 
further notations and results.

\section{Notation and definitions}

  The generic topologies considered in this work
are displayed in Fig. 1. Lines of the diagrams represent
scalar propagators $(p^2-m^2+i0)^{-1}$ with either 
mass $m$ or zero mass. 
For both ultraviolet and infrared divergences 
we use dimensional regularization
in the Minkowski space with dimension $D=4-2\varepsilon$.
Each loop integration is normalized as follows
\begin{equation}
\label{norm}
  \int d\tilde{k} = 
   (m^2 e^{\gamma_E})^\varepsilon \int \frac{d^Dk}{\pi^{D/2}}\,
\end{equation}
with $\gamma_E$ the Euler constant.

  Our starting point is the large mass expansion \cite{asymptotic} 
of the diagrams. Here we just note that the result of a large mass 
expansion for these diagrams reads ($z=q^2/m^2$)
\begin{equation}\label{lme}
  J = \frac{1}{(q^2)^a}\sum_{n\geq 1} z^n
       \sum_{j=0}^\omega \frac{1}{\varepsilon^j}
       \sum_{k=0}^\nu \log^k(-z)
                  A_{n,j,k}\,,
\end{equation}
where $a$ is the dimensionality of the diagram,
$\omega$ and $\nu$ independent of $n$ are
the highest degree of divergence and 
the highest power of $\log(-q^2/m^2)$, respectively
(in our cases $\omega,\nu\leq4$).
The coefficients $A_{n,j,k}$ are of the form
$r_1+\zeta_2 r_2+\dots+\zeta_\nu r_\nu$ with 
$r_c$ being rational numbers and $\zeta_c=\zeta(c)$ is the
Riemann $\zeta$-function.

  Series (\ref{lme}) always has a nonzero radius of
convergence, which is defined by the position
of the nearest nonzero threshold in the $q^2$-channel.
For brevity we shall call $m$-cuts ($2m$-cuts, etc.)
possible cuts of a diagram in $q^2$ corresponding
to 1 (2, etc.) massive particles in the intermediate state.

\section{Diagrams with $m$-cuts}

  We start from diagrams with the simplest threshold
structure, i.e. with $m$-cuts and (possible)
$0$-cuts. Such diagrams are particulary easy to
handle and results for all of them within the considered generic
topologies of Fig. 1 can be found in the appendices.
It will be shown that all are expressible in terms of
harmonic sums.

  As a first example consider the 2-loop 2-point function
of Fig. 2, $I_{13}$, which was considered in \cite{BroadhurstZP47}.
Using the standard large mass 
expansion technique one can get the first few coefficients of the
expansion of this diagram in powers of $z=q^2/m^2$
\begin{equation}\label{sample1}
  \frac{1}{q^2}\sum_{n=1}^{\infty} a_n z^n
  = \frac{1}{q^2}\Biggl(
         2\zeta_2 z
       + \left( \zeta_2 + \frac12 \right) z^2
       + \left( \frac23\zeta_2 + \frac12 \right) z^3
       + \left( \frac12\zeta_2 + \frac{65}{144} \right) z^4
%       + \left( \frac25\zeta_2 + \frac{29}{72}  \right) z^5
       + \dots \Biggr)\,,
\end{equation}
where $\zeta_2=\zeta(2)$ is the Riemann $\zeta$-function.

  It takes seconds to evaluate the first 10 terms 
(we use a package written in FORM \cite{FORM}) but the evaluation time
increases exponentially for the higher order terms.
Although 10--20 coefficients are sufficient to get a
very precise numerical value for the diagram, even at $|z|\sim 10^2$
(via conformal mapping and accelerated convergence \cite{ft}), this
cannot be called a satisfying result.

  Our key observation is that the information contained
in the first 10 (or even less!) coefficients of the series 
(\ref{sample1}) is enough to restore the whole series.
Namely, each coefficient $a_n$ can be expressed as some
combination of harmonic sums and their extensions.
In this case the series can be continued numericaly or analyticaly
to any point of the complex $z$-plane. 

  Thus our first step is to find an expression for
the higher order terms of the series (\ref{sample1}) as 
a functions of $n$ i.e. $a_n=a(n)$.

To achieve this let us search for $a_n$ as a linear
combination of elements of a certain kind. For these we take
$1/n$ and $S_k(n-1)=\sum_{j=1}^{n-1} 1/j^k$ (harmonic sum).
As a matter of observation, the rule is that one has to
take into account all possible products of the type 
$\zeta_a S_{b_1}\dots S_{b_k}/n^c$ with 
the 'transcendentality level' (TL) $a+b_1+\dots+b_k+c=3$. 
It is obvious that one
can exclude $\zeta_3$ beforehand since it never appears on the
r.h.s. of (\ref{sample1}).
Thus we have the following Ansatz for $a_n$
\begin{equation}\label{ansatz}
  a_n = \frac{\zeta_2}{n}x_1 + \zeta_2 S_1 x_2
       + S_3 x_3 + S_2 S_1 x_4 + S_1^3 x_5
       + \frac{S_2}{n} x_6 + \frac{S_1^2}{n} x_7
       + \frac{S_1}{n^2} x_8 + \frac{1}{n^3} x_9\,,
\end{equation}
where $x_1,\dots,x_9$ are rational numbers independent
on $n$ and we omitted the
arguments in the $S$-functions. These can be taken with
argument $n$ or $n-1$. 
We choose the latter option. However, either choice gives a solution
since the difference is only in rearranging $1/n$ terms.
Later on we refer to the structures in (\ref{ansatz})
as `basis elements' or just `basis'. Indeed, the functions
$\zeta_aS_{b_1}(n-1)\dots S_{b_k}(n-1)/n^c$ are
algebraically independent.

  Inserting the expression for $a_n$ from (\ref{ansatz}) 
into the l.h.s. of (\ref{sample1}) 
and equating equal powers of $z$, we obtain
a system of linear equations for the $x_i$.
One needs at least 9 first coefficients on
the r.h.s. of (\ref{sample1}) to solve the equations for 9
variables and any system of more than 9 equations should
be consistent and have the same solution if the Ansatz 
(\ref{ansatz}) is correct. 
An explicit computation ensures that the system can be solved
in terms of rational numbers for the $x_i$
and this latter consistency property holds.
The solution is $x_1=2,x_6=2,x_8=-2,x_{2,3,4,5,7,8}=0$ i.e.
the answer for the diagram at hand is 
\begin{equation}\label{res1}
I_{13} =
  \frac{1}{q^2} 
  \sum_{n=1}^{\infty} z^{n} 
       \Biggl(
     2 \frac{\zeta_2}{n} + 2 \frac{S_2(n-1)}{n} - 2\frac{S_1(n-1)}{n^2}
       \Biggr)\,.
\end{equation}

  This is the desired result. One notices that
in (\ref{sample1}) and Ansatz (\ref{ansatz})
the terms proportional to $\zeta_2$ are linearly
independent and can be treated separately,
i.e one could solve a system of linear equations
for $x_1,x_2$ and another one for $x_3,\dots,x_9$.
In this sence only the 7 first coefficients of the
large mass expansion are required.  

  Series (\ref{res1}) converges for $|z|<1$ and 
represents in the domain of convergence an analytic
function. It can be analytically continued into the
whole $z$-plane except for the cut on the real axis
starting at $z=1$ where $I_{13}$ has a branch
point (threshold). In particular for the given diagram, 
summing the series, we find the representation
\begin{equation}\label{int1}
I_{13} =
   \frac{1}{q^2} \Biggl(
     -2\zeta_2\log(1-z) - 2\log(1-z)\,{\rm Li}_2(z)
     - 6 S_{1,2}(z) \Biggr)\,, 
\end{equation}
where ${\rm Li}_\nu(z)$ is a polylogarithm and 
$S_{p,q}(z)$ are the generalized Nielsen polylogarithms \cite{Devoto}
(see also Appendix C).

  Both representations
(\ref{res1}) and (\ref{int1}) are equivalent in the
sense that they both represent the same analytic
function and can be unambiguously converted into
each other. However, we
find it more convenient to use the series representation
and refer  for the continuation to App. C and E.
Moreover a series representation is
usually shorter than a representation
in terms of known functions and/or an integral representation.

  Equation (\ref{ansatz}) can be considered as expansion of
the general coefficient $a_n$ in terms of a `basis' with 
(unknown, rational) coefficients $x_i$. Now the question arises:
what are the basis elements in general? 
Definitely there must be a certain connection
between the topology and the structure of thresholds of a diagram
on the one hand
and the structure of the basis elements on the other hand.
So far we are lacking rules for predicting a basis of a given diagram.
The power of the method, however, is that given a set of basis
elements for one diagram, it can be used to find the solution
for other diagrams. Often, though, one has to `generalize'
already known (harmonic) sums in a basis. In Sect. 4 we explain 
how to do this. 

  In particular it will be shown that
many 2-point and 3-point functions have similar bases
and thus solving some 2-point integrals (which are simpler)
we can find solutions for 3-point integrals. As an illustration
consider the diagram shown in Fig. 3, $P_5$. Its large mass expansion
looks like
\begin{equation}\label{sample2}
  P_5 =
   \frac{1}{(q^2)^2} \sum_{n=1}^\infty
     z^n \Biggl(
       r_n^{(2)} \log^2(-z)
     + r_n^{(1)} \log(-z) 
     + r_n^{(0,3)} \zeta_3 + r_n^{(0,2)} \zeta_2
     + r_n^{(0,0)}
                         \Biggr)\,,
\end{equation}
with the $r$'s being rational numbers. It is obvious that
one can search for a solution for each of the $r$'s independently.
Again we use the same set of functions ($1/n^a$ and $S_b$) as above
but now with different transcendentality level(s).
The system of equations has a solution only if we add
the factor $(-)^n$ which can be seen easily 
by inspection of the series. At the end we arrive at
\begin{eqnarray}\label{res2}
  P_5 \z=\z 
   \frac{1}{(q^2)^2} \sum_{n=1}^\infty
     z^n (-)^n \Biggl(
       \frac{S_1}{n} \log^2(-z)
     + \Bigl( - 4 \frac{S_2}{n} + \frac{S_1^2}{n} 
      - 2 \frac{S_1}{n^2} \Bigr) \log(-z) 
                    \nonumber\\
  \z\z  - 6 \frac{\zeta_3}{n} + 2 \frac{\zeta_2 S_1}{n}
     + 6 \frac{S_3}{n} - 2 \frac{S_2 S_1}{n} 
     + 4\frac{S_2}{n^2} - \frac{S_1^2}{n^2}
     + 2\frac{S_1}{n^3}
                         \Biggr)\,,
\end{eqnarray}
where as above we take all harmonic sums ($S_i's$) with the
argument $n-1$, which is omitted.\\

  The similarity of (\ref{res1}) and (\ref{res2}) is
spectacular. There are two comments in order.\\

At first we point out that in (\ref{res2})
the part without $\log(-z)$ has basis elements
$\zeta_aS_b/n^c$ obeying $a+b+c=4$ (i.e. it has a basis
with TL$=4$). Terms proportional to $\log(-z)$ are of 3rd level
while those proportional to $\log^2(-z)$ of the 2nd.
One can say that $\log(-z)$ itself is of 1st level
and each $\log^a(-z)$ 
reduces the level of basis elements by $a$ units.
This is the general behaviour for all diagrams we consider.
The same rule applies to UV and IR poles $1/\varepsilon$
if they are present in a diagram i.e. each $1/\varepsilon^a$
reduces the level of basis elements by $a$ units.

Secondly we wish to underline the presence of the
factor $(-)^n$. Strictly speaking the basis now is not
$\zeta_aS_b/n^c$ any more but rather $(-)^n\zeta_aS_b/n^c$.
In principle one can expect that a mixture of both
bases occurs and this is indeed the case for some diagrams.
Moreover, the $(-)$ may stand not only in front of $S_a$
but also inside the harmonic summation. Thus we are led to the 
alternating harmonic sums\footnote{These sums were used
in \cite{DIS,FL}.}
$K_a(n-1)=\sum_{j=1}^{n-1}(-)^{j+1}/j^a$.

  It will turn out that the results obtained above
indicate already a general property, i.e. in 2-loops
all the selfenergy functions have TL=3 and all vertices have TL=4
(at least for the diagrams under consideration).

\section{Generalization of basis elements}

  It was mentioned above that
we cannot directly predict the basis of a diagram.
Therefore we experiment with basis elements of
diagrams calculated before to match the expansion coefficients
of other ones.
Another extremely useful trick is that
we first establish the structure of the lowest
level in a diagram (i.e. the structure of the coefficients
of the highest pole in $\varepsilon$ or the highest
power of $\log(-z)$) and then try to reconstruct
terms of higher level. 
If, say, the $1/\varepsilon^k$ coefficients
are expressed symbolically in terms of basis elements $f_{(a)}$ with TL=$a$
then the $1/\varepsilon^{k-1}$ part is expressed in terms of
some ${f_{(a+1)}}$ etc. Therefore we need rules
of generalization, i.e. to find higher level elements from lower level ones.
These are discussed in this Section.

  We start from the following observation. Let $J$ be
a $D$-dimensional Feynman integral 
(properly normalized to be dimensionless)
depending on $z=q^2/m^2$.
It can be formally
written as multiple series of the hypergeometric type
\begin{equation}\label{feydia}
  J(z) = \sum_{j_1,j_2,\dots,j_s}
  \frac{\Gamma(\alpha_1\{j,D\})\dots\Gamma(\alpha_\nu\{j,D\})}
       {\Gamma(\beta_1\{j,D\})\dots\Gamma(\beta_\rho\{j,D\})} 
       \,z^{\gamma\{j,D\}} \,, 
\end{equation}
where $\Gamma$ is the Euler $\Gamma$-function.
Here the symbols $\alpha_i\{j,D\}$, $\beta_i\{j,D\}$ 
and $\gamma\{j,D\}$ stand for some linear combinations
of the summation indices $j_k$ and
the space-time dimension $D=4-2\varepsilon$.
The form (\ref{feydia}), though rather obvious by itself,
can be deduced e.g. from the
Feynman parameter representation or $\alpha$-representation.
From this point of view, 'to evaluate' diagram $J$ means
to bring multiple sums in (\ref{feydia}) to some 
'known' hypergeometric form of
series or at least to remove as many summations as possible.

  Usually, however, one is interested in the expansion
of a diagram around $\varepsilon=0$.
On expansion, (\ref{feydia}) develops some singularities
in the $z$-plane and possible poles $1/\varepsilon$.
The expansion of $\Gamma$-functions produces $\psi$-functions
or equivalently $S$-sums. Namely for $j$ integer
\begin{equation}
\label{gamma}
  \frac{\Gamma(j+a\varepsilon)}{\Gamma(1+a\varepsilon)} =
   (j-1)!
   \exp\left\{ \sum_{k=1}^\infty a^k\varepsilon^k\, S_k(j-1) \right\}\,,
\end{equation}  
with $S_k(n)=\sum_{s=1}^{n}1/s^k$. This shows that harmonic sums naturally
play a special role in our subject.

  Terms of lowest order are obtained by replacing 
the exponent in (\ref{gamma}) by unity and (\ref{feydia})
becomes a multiple sum of factorials. Then the coefficient
of $z^n$ is a finite sum over factorials. Terms of higher level
are obtained by expanding the exponent in (\ref{gamma}) in $\varepsilon$,
which produces $S$-functions in (\ref{feydia}).

This suggests the following recipe of generalization: if $f_{(a)}$
is found to be a multiple (finite) sum 
$f_{(a)}(n)=\sum_{\{j\}}^n c_{\{j\}}$, then higher level
objects are obtained either by multiplying $f_{(a)}$
by $S$-functions or by inserting $S$-functions under the summation
sign of $f_{(a)}$ to give 
\begin{eqnarray}\label{ss}
\z\z  f_{(a+1)}(n) = \sum_{\{j\}}^n c_{\{j\}} S_1(\{j\}),\nonumber \\
\z\z  f_{(a+2)}(n) = \sum_{\{j\}}^n c_{\{j\}} S_2(\{j\}),
   \qquad
      f_{(a+1+1)}(n) = \sum_{\{j\}}^n c_{\{j\}} S_1(\{j\}) S_1(\{j\}),
                         \nonumber\\
\z\z  \dots \quad  \mbox{etc.,}
\end{eqnarray}
where the arguments of $S_k$ are some linear combinations of
summation indices. The same applies of course to the function
$S_a$ itself giving $S_{a,b}$ etc.

  From (\ref{ss}) it is seen that any $f_{(a)}$
produces in general many functions $f_{(a+1)}$ and higher ones.
In the case
of a multiple summation the number of possible combinations
of $S_k(\{j\})$ may get too large
and this general rule is getting rather formal than practical.
However, in practice we always met the situation that
$f_{a}$ is a simple (one-fold) sum.
Then, inserting $S_k(j-1)$ and $S_k(j)$ 
or equivalently $S_k(j-1)$ and $1/j$, 
(\ref{ss}) becomes
\begin{eqnarray}\label{ss1}
\z\z  f_{(a+1)}(n) = \sum_{j=1}^n c_j 
       \left\{ \frac{1}{j}, S_1(j-1) \right\},\nonumber \\
\z\z  f_{(a+2)}(n) = \sum_{j=1}^n c_j 
       \left\{ \frac{1}{j^2}, \frac{S_1(j-1)}{j}, 
            S_1^2(j-1), S_2(j-1) \right\},
                         \nonumber\\
\z\z  \dots \quad  \mbox{etc.}
\end{eqnarray}
The choice (\ref{ss1}) is justified {\it a posteriory}.

  This prescription, however,
must be modified if a diagram possesses $2m$-cuts 
(with a threshold at $q^2=4m^2$). Such diagrams produce
in the sum (\ref{feydia}) functions like 
$\Gamma(2j+a\varepsilon)$ and according to (\ref{gamma})
$S_k(2j-1)$. In such cases, along with $S_k(j-1)$, we insert
also $S_k(2j-1)$. 

  In conclusion of this Section we want to stress that
the recipes formulated above do not necessarily work
in every case for any diagram. But in many cases they do.
A possible reason of a failure of the 
prescription (\ref{ss}) may be e.g. the following : some
new basis element begins to contribute only, say, at 
the order $1/\varepsilon$ while at the order
$1/\varepsilon^2$ it happens to drop out. 
In such cases it will be missed in the analysis and
one has to use other more powerful and direct methods
of determination of low level terms.

%%%%%%%%%%%%%%%%%%%%%%%%%%%%%%%%%%%%%%%%%%%%%%%%%%%%%%%%%%%%%
%%%%%%%%%%%%%%%%%%%%%%%%%%%%%%%%%%%%%%%%%%%%%%%%%%%%%%%%%%%%%
%
%      W - functions
%
%%%%%%%%%%%%%%%%%%%%%%%%%%%%%%%%%%%%%%%%%%%%%%%%%%%%%%%%%%%%%
%%%%%%%%%%%%%%%%%%%%%%%%%%%%%%%%%%%%%%%%%%%%%%%%%%%%%%%%%%%%%

\section{Diagrams having both $m$- and $2m$-cuts and
$W$-functions}
  
  We now establish the structure of the lowest level
for the set of diagrams having both $m$- and $2m$-cuts.
While for diagrams with only $m$-cuts the complete
basis consists of elements $1/n^a,S_b,K_c$ and their
generalization due to (\ref{ss1}),
in the presence of $2m$-cuts the basis changes
drastically. In this case the behaviour of the coefficients
is governed by a new class of elements
which we call $W$. The generic structure
of the lowest level is
\begin{equation}\label{w1}
   W_1(n) = \sum_{j=1}^n {2j\choose j} \frac{1}{j}
\end{equation}
and higher level functions are obtained following
the rules of the previous Section.

  The $W$-structure can be deduced from different diagrams 
with $2m$-cuts
but most easily we find it from the nonplanar graph shown in
Fig. 3, $N_{12}$. Namely this diagram has 
a double collinear pole and we 
can easily perform both integrations in calculating the 
$1/\varepsilon^2$ contribution.

  Consider at first the massless lines 3 and 5 (see (\ref{norm}) 
for the normalization)
\begin{equation}
  N_{12} = 
  \int\frac{d\tilde{k}_1}{k_1^2 (k_1-p_1)^2} \phi(k_1,k_2,p_1,p_2),
\end{equation}
where $\phi$ stands for the rest of the diagram. Introducing the
Feynman parameter $\alpha_1$ this can be written as
\begin{equation}
  N_{12} = 
  \int_0^1 d\alpha_1 
  \int\frac{d\tilde{k}_1}{(k_1^2-2\alpha_1 k_1 p_1)^2} \phi
  =  \int_0^1 d\alpha_1 
  \int\frac{d\tilde{k}_1}{(k_1-\alpha_1 p_1)^4} \phi ,
\end{equation}
where we have used $p_1^2=0$. This integral diverges
in 4-dimensions when $(k_1-\alpha_1  p_1) \to 0$, i.e.
there is a collinear divergence. For the calculation of
the most singular term
in $\varepsilon$ one can effectively replace 
\begin{equation}
  \frac{1}{(k_1-\alpha_1 p_1)^4} \longrightarrow 
      i \frac{\pi^2}{\varepsilon} \delta^{(D)}(k_1-\alpha_1 p_1),
\end{equation}
and the same replacement applies to the pair of lines 4 and 6 
and integration $d\tilde{k_2}$. Therefore
\begin{equation}
  N_{12} \stackrel{1/\varepsilon^2}{=} 
  - \frac{\pi^4}{\varepsilon^2} \int_0^1 d\alpha_1 \int_0^1 d\alpha_2
   \int 
   \frac{
    d\tilde{k}_1\, d\tilde{k}_2\,
    \delta^{(D)}(k_1-\alpha_1 p_1)\,
    \delta^{(D)}(k_2-\alpha_2 p_2)
        }
        {[(k_1-k_2-p_1)^2 - m^2] [(k_2-k_1-p_2)^2 - m^2 ] }.
\end{equation}

  Integrating out $k_1,k_2$ with the help of the $\delta$-functions
and using $p_1^2=p_2^2=0$, $p_1p_2=q^2/2$,
we arrive at ($z=q^2/m^2$)
\begin{eqnarray}
\label{polepart}
N_{12} \z \stackrel{1/\varepsilon^2}{=} \z
  - \frac{1}{\varepsilon^2 (q^2)^2}\int_0^1
  \frac{d\alpha_1\, d\alpha_2}
  {(1-\alpha_1(1-\alpha_2)z) (1-(1-\alpha_1)\alpha_2 z)} \nonumber\\
  \z=\z
   -\frac{1}{\varepsilon^2 (q^2)^2}
    \sum_{n=1}^\infty z^n \frac{(n!)^2}{(2n-1)!}
    3\sum_{j=1}^{n-1} \frac{(2j)!}{(j!)^2} \frac{1}{j} \nonumber\\
  \z=\z
   -\frac{1}{\varepsilon^2 (q^2)^2}
    \sum_{n=1}^\infty z^n {2n\choose n}^{\!\!\!-1} 6\,\frac{W_1(n-1)}{n}.
\end{eqnarray} 

If we assign to 
the factor ${2n\choose n}^{\!\!\!-1}$ the 0th level,
then the expression on the r.h.s of (\ref{polepart}) has
the correct 2nd level as it should be for the double
pole part of a vertex function. Using the results of Sect. 4
we obtain functions of higher levels $W_2,W_{1,1},W_3$ etc.
(see Appendix A). 

%%%%%%%%%%%%%%%%%%%%%%%%%%%%%%%%%%%%%%%%%%%%%%%%%%%%%%%%%%%%%
%%%%%%%%%%%%%%%%%%%%%%%%%%%%%%%%%%%%%%%%%%%%%%%%%%%%%%%%%%%%%
%
%    Differential Equation Method
%
%%%%%%%%%%%%%%%%%%%%%%%%%%%%%%%%%%%%%%%%%%%%%%%%%%%%%%%%%%%%%
%%%%%%%%%%%%%%%%%%%%%%%%%%%%%%%%%%%%%%%%%%%%%%%%%%%%%%%%%%%%%

\section{Differential equation method}

   It has been demonstrated above that the most important
point in our approach to set up a Taylor series for the diagrams
under consideration is to find the proper basis elements. In the last
Section we have seen a special calculation to find one more basis 
element. There is no general way of how to find further such elements
but another effective approach is that of the differential equation
method (DEM) \cite{DEM}. It was used in \cite{FKV} to find the expansion
of diagrams $P_{126}$ and $P_{56}$ in the notation of Fig.3. 
Therefore we give in this Section a short review of the DEM and demonstrate
the finding of another basis element by calculating the two-point 
diagram $I_{15}$ of Fig. 2.

The DEM allows one to calculate massive diagrams by
reducing them to others having an essentially simpler structure 
\footnote{A similar technique has been introduced recently
by Remiddi \cite{Remiddi}.}.
More precisely, one writes a linear differential 
equation (w.r.t. masses) for the diagram under consideration.
The inhomogeneous term of the equation is a sum of
diagrams with smaller number of lines.
These in turn can be reduced via another differential
equation to even simpler diagrams and so on.
Solving the so obtained equations iteratively
one gets for the initial diagram an $s$-fold
integral representation which is in general simpler than
the usual parametric or dispersion integrals.
Often some of the integrations can be performed analytically.

  For completeness we give some formulae that 
will be useful in the futher discussion.
We introduce the following graphical representation
for the propagators
%
%  PROPAGATORS
%   
\vspace{3mm}\hfill\\
\hspace*{3cm}
$\frac{\displaystyle 1}{\displaystyle (q^2+i0)^\alpha}\,=\;\;$
\begin{picture}(30,10)(5,4)
\DashLine(0,5)(30,5){2}
\Vertex(0,5){1}
\Vertex(30,5){1}
\Text(15,7)[b]{$\scriptstyle\alpha$} 
\end{picture} 
$,\qquad \frac{\displaystyle 1}{\displaystyle (q^2-m^2+i0)^\alpha}\,=\;\;$
\begin{picture}(30,10)(5,4)
\Line(0,5)(30,5)
\Vertex(0,5){1}
\Vertex(30,5){1}
\Text(15,7)[b]{$\scriptstyle\alpha$} 
\Text(15,3)[t]{$\scriptstyle m^2$} 
\end{picture} 
\hfill (19)\\ 
\vspace{3mm}\hfill\\
$\alpha$ and $m$ are called the index and mass of this line.
Further (except when mentioned otherwise) all solid lines have the 
same mass $m$. Lines with index 1 and mass $m$ are not marked.

  The basic tool is the integration by part relation \cite{bypart}
which is obtained by multiplying the integrand of a diagram
with $(d/dk_\mu) k_\mu$ and using $\int d^D {\rm div}(\dots)=0$.
As a result for a triangle 
\footnote{Similar relations may easily be obtained for arbitrary $n$-point
diagrams (see e.g. \cite{DEM} and discussion in \cite{bubbl}).}
with arbitrary masses one has
(here lines with index $\alpha_i$ have mass $m_i$)
%
% TRIANGLE RULE
%  in the correct metrics!
%
\vspace{3mm} \hfill \\
\hspace*{1cm}
\trg{\alpha_1}{\alpha_2}{\alpha_3}
$(D-2\alpha_1-\alpha_2-\alpha_3) \,=\, 2 m_1^2\alpha_1\,\,\,$
\trg{\alpha_1+1}{\alpha_2}{\alpha_3}
\vspace{20pt} \hfill \\
\vspace{-0pt}
\vspace{-20pt}\hfill \\
\hspace*{1cm}
$+ \Biggl\{ \alpha_2\Biggl(\;\;$
\trg{\alpha_1-1}{\alpha_2+1}{\alpha_3}
$+\;\;\;\;\;$
\trgleft{\alpha_1}{\alpha_2+1}{\alpha_3}
$\,+(m_1^2+m_2^2)\;\;\;\;$
\trg{\alpha_1}{\alpha_2+1}{\alpha_3}
$\Biggr) + (\alpha_2\leftrightarrow\alpha_3) \Biggr\}$
\hfill (20)\\ 
\vspace{3mm} \hfill \\
We stress the fact that the basic line of the triangle plays a
special role (in the following we call it 'distinctive').
This relation is used in the following manner. All triangles
on the r.h.s. with 'shifted' indicies $\alpha_i+1$ can be
written as derivatives of the initial triangle w.r.t. $m^2_i$.
Generally speaking, one has to consider all masses $m_i$ to
be different and put them equal at the end of the calculation.
In practice one can often eliminate some of the derivatives
by combining several differential equations.

  Whenever a self-energy subdiagram occurs the following
relation is used
%
% SELF-ENERGY
%
\vspace{3mm}\hfill\\
\hspace*{0.1cm}
\begin{picture}(60,30)(0,13)
\Curve{(5,15)(30,25)(55,15)}
\Curve{(5,15)(30,5)(55,15)}
\Line(5,15)(-5,15)
\Line(55,15)(65,15)
\Text(30,27)[b]{$\scriptstyle\alpha_1,\,m_1$} 
\Text(30,3)[t]{$\scriptstyle\alpha_2,\,m_2$} 
\Text(0,12)[t]{$q$} 
\end{picture} 
$\displaystyle   \;\,=\,
   i^{1+D}
   \frac{\Gamma(\alpha_1+\alpha_2-D/2)}{\Gamma(\alpha_1)\,\Gamma(\alpha_2)}
   \int_0^1 \frac{ds}{(1-s)^{\alpha_1+1-D/2}s^{\alpha_2+1-D/2}}\;\;\;$
\begin{picture}(70,30)(0,4)
\Line(5,5)(65,5)
\Vertex(5,5){1}
\Vertex(65,5){1}
\Line(5,5)(-5,5)
\Line(65,5)(75,5)
\Text(33,7)[b]{$\scriptstyle\alpha_1+\alpha_2-D/2$} 
\Text(33,3)[t]{$\scriptstyle \frac{m_1^2}{1-s}+\frac{m_2^2}{s}$} 
\Text(-3,-2)[b]{$q$} 
\end{picture}
\hfill (21)\\ 
\vspace{3mm}\\
This allows one to reduce an $l$-loop 
to an $(l-1)$-loop diagram with one propagator
having 'mass' $m_1^2/(1-s)+m_2^2/s$.

  We now apply this technique to the evaluation of
diagrams with one mass. Consider first the symmetric planar
diagram $P_{126}$ (Fig. 3). 
It possesses only $2m$-cuts and zero-cuts
but no $m$-cuts. Therefore the nearest singularity
in the complex $z=q^2/m^2$ plane is placed
at $z=4$ and the diagram obviously has some new
structures different from those already discussed.
Applying (20) with distinctive central rung we
have
%
%  Case P126
%
\vspace{-6mm}\hfill\\
\hspace*{0cm}
$(D-4)\,$
\begin{picture}(35,60)(5,27)
\DashLine(5,55)(35,55){2}
\DashLine(5,30)(5,55){2}
\DashLine(35,30)(35,55){2}
\Line(5,30)(35,30)
\Line(20,5)(5,30)
\Line(20,5)(35,30)
\Line(20,5)(20,3)
\Line(5,55)(3,58)
\Line(35,55)(38,58)
\end{picture}
$\,=\, 2\!\!$
\begin{picture}(35,60)(5,27)
\DashLine(5,55)(35,55){2}
\DashLine(20,30)(5,55){2}
\DashLine(20,30)(35,55){2}
\BCirc(20,17.5){12.5}
\Line(20,5)(20,3)
\Line(5,55)(3,58)
\Line(35,55)(38,58)
\Text(35,15)[l]{$\scriptstyle 2$}
\end{picture}
$\!\! -\,2\;$
\begin{picture}(35,60)(5,27)
\DashLine(5,55)(35,55){2}
\DashLine(5,30)(5,55){2}
\Line(5,30)(35,55)
\Line(20,5)(5,30)
\Line(20,5)(35,55)
\Line(20,5)(20,3)
\Line(5,55)(3,58)
\Line(35,55)(38,58)
\Text(25,15)[l]{$\scriptstyle 2$}
\end{picture}
$ -\,4m^2\,$
\begin{picture}(35,60)(5,27)
\DashLine(5,55)(35,55){2}
\DashLine(5,30)(5,55){2}
\DashLine(35,30)(35,55){2}
\Line(5,30)(35,30)
\Line(20,5)(5,30)
\Line(20,5)(35,30)
\Line(20,5)(20,3)
\Line(5,55)(3,58)
\Line(35,55)(38,58)
\Text(29,15)[l]{$\scriptstyle 2$}
\end{picture}
$ -\,2m^2\,$
\begin{picture}(35,60)(5,27)
\DashLine(5,55)(35,55){2}
\DashLine(5,30)(5,55){2}
\DashLine(35,30)(35,55){2}
\Line(5,30)(35,30)
\Line(20,5)(5,30)
\Line(20,5)(35,30)
\Line(20,5)(20,3)
\Line(5,55)(3,58)
\Line(35,55)(38,58)
\Text(20,34)[l]{$\scriptstyle 2$}
\end{picture}
\hfill (22)\\ 
\vspace{5mm} \hfill\\
It is easy to see that
the last two terms on the r.h.s. can be combined into the total
derivative of the initial diagram w.r.t. $m^2$. 
Therefore we immediately obtain the differential
equation for our diagram. The inhomogeneous term now
is the sum of the first and the second diagram on the r.h.s. of (22).
The former is trivial while for the latter we repeat the
procedure with the left vertical line being distinctive:
%
%  second equation for case P126
%
\vspace{-6mm}\hfill\\
\hspace*{0mm}
$(D-4)$
\begin{picture}(35,60)(5,27)
\DashLine(5,55)(35,55){2}
\DashLine(5,30)(5,55){2}
\Line(5,30)(35,55)
\Line(20,5)(5,30)
\Line(20,5)(35,55)
\Line(20,5)(20,3)
\Line(5,55)(3,58)
\Line(35,55)(38,58)
\Text(25,15)[l]{$\scriptstyle 2$}
\end{picture}
$=$
\begin{picture}(35,60)(5,23)
\Line(20,5)(5,40)
\Line(20,5)(35,40)
\Curve{(5,40)(20,33)(35,40)}
\DashCurve{(5,40)(20,47)(35,40)}{2}
\Line(20,5)(20,3)
\Line(5,40)(3,43)
\Line(35,40)(38,43)
\Text(28,15)[l]{$\scriptstyle 2$}
\Text(20,49)[b]{$\scriptstyle 2$}
\end{picture}
$+$
\begin{picture}(35,60)(5,23)
\Line(20,5)(5,40)
\Line(20,5)(35,40)
\Curve{(5,40)(20,33)(35,40)}
\DashCurve{(5,40)(20,47)(35,40)}{2}
\Line(20,5)(20,3)
\Line(5,40)(3,43)
\Line(35,40)(38,43)
\Text(28,15)[l]{$\scriptstyle 2$}
\Text(20,31)[t]{$\scriptstyle 2$}
\end{picture}
$-$
\begin{picture}(35,60)(5,23)
\DashLine(20,5)(5,40){2}
\DashLine(5,40)(35,40){2}
\Curve{(20,5)(25,27)(30,36)(35,40)}
\Curve{(20,5)(25,9)(30,17)(35,40)}
\Line(20,5)(20,3)
\Line(5,40)(3,43)
\Line(35,40)(38,43)
\Text(22,25)[r]{$\scriptstyle 2$}
\Text(32,15)[l]{$\scriptstyle 2$}
\end{picture}
\hfill (23)\\
\vspace{3mm} \hfill\\
The resulting diagrams on the r.h.s. all have self-energy
insertions which can be replaced according to (21)
and the remaining triangles are solved by usual Feynman
parameters.

  After some rather long but trivial transformations
we obtain 
\setcounter{equation}{23}
\begin{eqnarray}
P_{126}  \z=\z -\frac{1}{2(q^2)^2}
  \int^1_0 \frac{ds}{s(1-s)} 
   \Biggl[ \frac{1}{\varepsilon^2} \log(1-\xi) 
   + \frac{1}{\varepsilon} \biggl( \frac12\log^2(1-\xi) 
            - \log(-z)\log(1-\xi) \biggr) 
              \nonumber \\
   \z+\z \log[s(1-s)]\log^2(1-\xi) - \frac76\log^3(1-\xi)
   - 12 S_{1,2}(\xi) - 5 \log(1-\xi){\rm Li}_2(\xi) 
    \nonumber \\
  \z+\z \frac12\log(-z)\log^2(1-\xi) 
   + \frac12\log^2(-z)\log(1-\xi)
   \Biggr],
\end{eqnarray}
were $\xi=zs(1-s)$ and $S_{1,2}(z)$ is the Nielsen polylogarithm
(see Appendix C). The above remaining integral is very difficult
to evaluate analytically. However the expansion in $z$ 
is welcome
\begin{eqnarray}
P_{126} \z=\z  
   \frac{1}{(q^2)^2}
    \sum_{n=1}^{\infty} z^n {2n\choose n}^{\!\!\!-1}\frac{1}{n^2}
   \Biggl\{
      \frac{1}{\varepsilon^2}
    + \frac{1}{\varepsilon} \Bigl[ 
           - S_1
           - \log(-z)
             \Bigr] \nonumber\\
 \z\z
      - \frac32 S_2
      - \frac{15}{2} S_1^2
      + 4 S_1 \overline{S}_1
      + 2 \frac{S_1}{n}
      - S_1 \log(-z)
      + \frac12 \log^2(-z)
      \Biggr\},
\end{eqnarray}
with $S_a=S_a(n-1)$ and $\overline{S}_1=S_1(2n-1)$. This
expression has the correct threshold at $q^2=4m^2$ due
to the factor ${2n\choose n}^{\!\!\!-1}$. The appearence
of $\overline{S}$ is somehow expected as was mentioned
at the end of Sect. 5, but the factor 
${2n\choose n}^{\!\!\!-1}$ is difficult to guess without
additional information (compare however with (\ref{polepart}) ).

  We now turn to the 2-point asymmetric diagram 
$I_{15}$ (Fig. 2) which contains
another new basis element.
Using (20) with distinctive
central rung for the left triangle we obtain the following
differential equation
%
%  First equation
%
\vspace{0mm} \hfill \\
$\Biggl[ D-4-2m^2\frac{\displaystyle d}{\displaystyle dm^2} \Biggr]$
\hspace*{2mm}
\begin{picture}(60,30)(5,13)
\Line(0,15)(30,30)
\DashLine(0,15)(30,00){2}
\DashLine(60,15)(30,30){2}
\DashLine(60,15)(30,00){2}
\Line(30,00)(30,30)
\Line(0,15)(-5,15)
\Line(60,15)(65,15)
\end{picture}
$\,\,=\,\,$
\hspace*{2mm}
\begin{picture}(60,30)(5,13)
\CArc(15,15)(15,0,180)
\DashCArc(15,15)(15,180,360){2}
\DashCArc(45,15)(15,0,180){2}
\DashCArc(45,15)(15,180,360){2}
\Line(0,15)(-5,15)
\Line(60,15)(65,15)
\Text(15,35)[]{$\scriptstyle 2$}
\end{picture}
$\,\,+\,\,$
\hspace*{2mm}
\begin{picture}(60,30)(5,13)
\CArc(15,15)(15,0,180)
\DashCArc(15,15)(15,180,360){2}
\DashCArc(45,15)(15,0,180){2}
\DashCArc(45,15)(15,180,360){2}
\Line(0,15)(-5,15)
\Line(60,15)(65,15)
\Text(15,5)[]{$\scriptstyle 2$}
\end{picture}
\\
\hspace*{10mm}
$+\,\,$
\hspace*{2mm}
\begin{picture}(45,30)(5,13)
\Line(0,15)(30,30)
\DashLine(0,15)(30,00){2}
\DashCArc(30,15)(15,-90,90){2}
\Line(30,00)(30,30)
\Line(0,15)(-5,15)
\Line(30,30)(45,30)
\Text(15,28)[]{$\scriptstyle 2$}
\end{picture}
$\,\,+\,\,$
\hspace*{2mm}
\begin{picture}(45,30)(5,13)
\Line(0,15)(30,30)
\DashLine(0,15)(30,00){2}
\DashCArc(30,15)(15,-90,90){2}
\Line(30,00)(30,30)
\Line(0,15)(-5,15)
\Line(30,00)(45,00)
\Text(15,3)[]{$\scriptstyle 2$}
\end{picture}
$+\,m^2$
\hspace*{2mm}
\begin{picture}(60,30)(5,13)
\Line(0,15)(30,30)
\DashLine(0,15)(30,00){2}
\DashLine(60,15)(30,30){2}
\DashLine(60,15)(30,00){2}
\Line(30,00)(30,30)
\Line(0,15)(-5,15)
\Line(60,15)(65,15)
\Text(15,3)[]{$\scriptstyle 2$}
\end{picture}
\vspace{7mm} \hfill \\
  One gains little from this because of the last term
on the r.h.s. This has the structure of the initial diagram 
but with one massless propagator squared. 
To get rid of this term
we can write another equation,
using the second massive line as distinctive of the triangle:
%
%  Second equation
%
\vspace{0mm} \hfill \\
$\Biggl[ D-4-2m^2\frac{\displaystyle d}{\displaystyle dm^2} \Biggr]$
\hspace*{2mm}
\begin{picture}(60,30)(5,13)
\Line(0,15)(30,30)
\DashLine(0,15)(30,00){2}
\DashLine(60,15)(30,30){2}
\DashLine(60,15)(30,00){2}
\Line(30,00)(30,30)
\Line(0,15)(-5,15)
\Line(60,15)(65,15)
\end{picture}
$\,\,=\,\,$
\hspace*{-4mm}
\begin{picture}(60,30)(5,13)
\DashCArc(30,15)(15,90,270){2}
\DashLine(60,15)(30,30){2}
\DashLine(60,15)(30,00){2}
\Line(30,00)(30,30)
\Line(60,15)(65,15)
\Line(30,30)(15,30)
\Text(33,15)[]{$\scriptstyle 2$}
\end{picture}
$\,\,+\,\,$
\hspace*{-4mm}
\begin{picture}(60,30)(5,13)
\DashCArc(30,15)(15,90,270){2}
\DashLine(60,15)(30,30){2}
\DashLine(60,15)(30,00){2}
\Line(30,00)(30,30)
\Line(60,15)(65,15)
\Line(30,30)(15,30)
\Text(18,15)[]{$\scriptstyle 2$}
\end{picture}
\\
\hspace*{10mm}
$+\,\,$
\hspace*{2mm}
\begin{picture}(45,30)(5,13)
\Line(0,15)(30,30)
\DashLine(0,15)(30,00){2}
\DashCArc(30,15)(15,-90,90){2}
\Line(30,00)(30,30)
\Line(0,15)(-5,15)
\Line(30,30)(45,30)
\Text(27,15)[]{$\scriptstyle 2$}
\end{picture}
$+\,(q^2+m^2)$
\hspace*{2mm}
\begin{picture}(60,30)(5,13)
\Line(0,15)(30,30)
\DashLine(0,15)(30,00){2}
\DashLine(60,15)(30,30){2}
\DashLine(60,15)(30,00){2}
\Line(30,00)(30,30)
\Line(0,15)(-5,15)
\Line(60,15)(65,15)
\Text(15,3)[]{$\scriptstyle 2$}
\end{picture}
\vspace{7mm} \hfill \\
Taking a proper linear combination of these equations, we can
exclude the bad last term. In the resulting equation
the inhomogeneous term consists only of diagrams with self-energy
insertions which can be solved immediately. As a result 
for $I_{15}$ we write its expansion in $z$
\begin{equation}
\label{i15}
I_{15} =
  \frac{1}{q^2} 
  \sum_{n=1}^{\infty} z^n \frac{1}{n}
       \Biggl[
     - \frac{1}{n}\log(-z) - \zeta_2
     + 3 \sum_{j=1}^{n-1}{2j\choose j}^{\!\!\!-1}\frac{1}{j^2}
     + \frac{2}{n^2} + 4 {2n\choose n}^{\!\!\!-1} \frac{1}{n^2}
       \Biggr]. 
\end{equation}

   The inner sum in (\ref{i15}) is the new type of basis elements
which we discovered from this calculation. In a somewhat more 
general way of writing it is
\begin{equation}
  V_a(n-1) = \sum_{j=1}^{n-1} {2j\choose j}^{\!\!\!-1} \frac{1}{j^a}.
\end{equation} 

   Further basis elements were obtained in a similar fashion. They
are listed in the appendix. We do not give all the details of their
derivation, which was done in a way analogous to the above.

%%%%%%%%%%%%%%%%%%%%%%%%%%%%%%%%%%%%%%%%%%%%%%%%%%%%%%%
%%%%%%%%%%%%%%%%%%%%%%%%%%%%%%%%%%%%%%%%%%%%%%%%%%%%%%%
%
%  Continuation
%
%%%%%%%%%%%%%%%%%%%%%%%%%%%%%%%%%%%%%%%%%%%%%%%%%%%%%%%
%%%%%%%%%%%%%%%%%%%%%%%%%%%%%%%%%%%%%%%%%%%%%%%%%%%%%%%

\section{Analytical continuation}

   Taking care of the zero-thresholds by factorizing $\log(-z),( z=q^2/M^2 )$
as described above,
all series under consideration have finite radius of
convergence $|z|<|z_0|$ where $z_0$ is the lowest
non-zero (pseudo)-threshold.
Therefore these series need to be analytically continued
on their cut since this is in general the region of
physical interest.
The most straigthforward way is the continuation
via conformal mapping and special summation techniques
(usually Pad\'{e} approximants) \cite{ft}. In principle this 
allows one to
continue the diagrams to the whole complex $z$-plane including
the cut; at least this method works perfectly for low
and moderate $|z/z_0|$. For the diagrams we consider here, the
situation is even better than in cases with several masses since
our analytic results for the Taylor coefficients
allow to calculate arbitrary many of them easily
with any desired precision.\\

   Nevertheless it is of interest to investigate the domain
$|z/z_0|\gg1$ and 
have series in powers of $1/z$. The first few coeficients
of such series can of course be obtained by expanding any diagram
in the `standard' manner \cite{asymptotic}, \cite{propagator},
\cite{Smirnov} in large $q^2$. 
This expansion is known
to be much more difficult than the corresponding
$q^2/M^2$-expansion. One of the reasons is that the number of contributing
subgraphs in this case is usually larger.
In the language of the present paper the `basis' is more extended
(see examples below).\\

   Here we present a different approach for the diagrams with
one non-zero mass, starting from the series in $z=q^2/M^2$.
The idea is to apply the Sommerfeld--Watson transformation 
to the $z$-series : given $J=\sum_{n=1}^\infty c_n z^n$,
it can be written as a Sommerfeld--Watson contour integral
\begin{equation}
\label{SW}
   J = \frac{1}{2\pi i} 
     \int\limits_{C_1+C_2} \frac{e^{i\pi n}\,c_n z^n}{\sin(\pi n)}\, dn.
\end{equation}
The integral in (\ref{SW}) is to be performed along a closed contour in
the $n$-plane consisting of two pieces: $C_1$ is a line parallel
to the imaginary axis $(\gamma-i\infty,\gamma+i\infty)$ with
$0<\gamma<1$ and $C_2$ is a semicircle closed in the 
right-half-plane at infinity.

  The representation (\ref{SW}) is valid of course only if the
coefficients $c_n$ (considered as analytic function of $n$)
decrease fast enough at infinity and
do not have singularities to the right
of $C_1$. Moving $C_1$ to the left in the negative direction,
one has to add the corresponding residues. We now suppose that
$c_n$ is regular
in the whole $n$-plane except (possible) poles at nonpositive
integers $n=-k$, $k=0,1,2,\dots$ and has the required
behaviour at infinity. Then

\begin{equation}
   J = - \sum_{k=0}^{\infty}
       {\rm res}_{n=-k}  \frac{e^{i\pi n}\,c_n z^n}{\sin(\pi n)}.
\end{equation}

   The crucial point now is to show how to continue analytically 
each of the structures occuring in this paper. 
First let us note without proof
that all singularities of these functions are placed
on the real axis at nonpositive integers. Therefore
we need to continue them through the neigborhood
of these points i.e $n=-k+\delta$ with $k=0,1,2,\dots$
and $|\delta|<1$.

  To start with, let us consider one-fold sums. First of all we
note that $S_1$ is continued by
\begin{equation}
\label{S1}
S_1(n-1) = \psi(n)+\gamma_E\,
\end{equation}
where $\psi(n)$ is the logarithmic derivative of the $\Gamma$-function.
All sums $S_a$ with $a>1$ are continued via the $\psi$-function and
its derivatives. All $K_a$ are simply expressed through $S_a$
\begin{equation}
K_a(n) = S_a(n) - 2^{1-a} S_a([n/2]).
\end{equation}
To continue the other functions we use the following general
method. Let $f(n-1)=\sum_{j=1}^{n-1}\phi(j)$. Then, formally
\begin{eqnarray}
\label{diff}
  f(n-1) = \sum_{j=1}^{\infty}\phi(j)
             - \sum_{j=n}^{\infty}\phi(j) 
         = f(\infty)
             - \sum_{j=0}^{\infty}\phi(j+n).
\end{eqnarray}
By inspection of the elements under consideration
we see that $\phi(n)$ is always an analytic function
in the whole $n$-plane apart from (possible) poles at
nonpositive integers. Then, with $n=-k+\delta$,
we obtain for the last term in (\ref{diff}) 

\begin{eqnarray}
\label{split}
\sum_{j=0}^{\infty}\phi(j-k+\delta) \z=\z
   \sum_{j=0}^{k-1}\phi(j-k+\delta) +  \phi(\delta) 
          + \sum_{j=k+1}^{\infty}\phi(j-k+\delta) \nonumber\\
   \z=\z
   \sum_{j=1}^{k}\phi(-j+\delta) + \phi(\delta)
          + \sum_{j=1}^{\infty}\phi(j+\delta).
\end{eqnarray}
It is convenient to introduce the notation 
$f^{(\delta)}(n-1)=\sum_{j=1}^{n-1}\phi(j+\delta)$ for
the sum with 'shifted' summation index. 
Combining (\ref{diff}) and (\ref{split}) we finaly get
\begin{equation}
 f(-k-1+\delta) = f(\infty) - f^{(\delta)}(\infty)
   - \sum_{j=1}^k \phi(-j+\delta) - \phi(\delta),
\end{equation}
with $k=0,1,2,\dots$. The remaining finite sum can be
easily transformed in our cases to a 'known' function
and (possible) poles in $\delta$ can be explicitely
separated. In this way we obtain the following folmulae 
\begin{eqnarray}
S_a(-k-1+\delta) \z=\z 
  S_a(\infty) 
  - S_a^{(\delta)}(\infty)
  - (-)^a S_a^{(-\delta)}(k)
  - \frac{1}{\delta^a},         \label{Sa}   \\
K_a(-k-1+\delta) \z=\z 
  K_a(\infty) 
  - K_a^{(\delta)}(\infty)
  - (-)^{a+2\delta} K_a^{(-\delta)}(k)
  + \frac{(-1)^\delta}{\delta^a},    \\
V_a(-k-1+\delta) \z=\z 
  V_a(\infty) 
  - V_a^{(\delta)}(\infty)
  + (-)^a \pi{\rm ctg}(\pi\delta) \, W_{a-1}^{(-\delta)}(k)
  - {2\delta\choose\delta}^{\!\!\!-1}\frac{1}{\delta^a},  \label{Va} \\
W_a(-k-1+\delta) \z=\z 
  W_a(\infty) 
  - W_a^{(\delta)}(\infty)
  + (-)^a \frac{1}{\pi{\rm ctg}(\pi\delta)}\, V_{a+1}^{(-\delta)}(k)
  - {2\delta\choose\delta}\frac{1}{\delta^a}.  \label{Wa}
\end{eqnarray}
 The continuation of double etc. sums is done in a similar manner. 
In the inner summation  one uses the continuation 
of $S_a$ via $\psi$-function or (\ref{Sa}). The results are similar
to (\ref{Sa})-(\ref{Wa}) e.g.
\begin{eqnarray}
\z\z
V_{a,1}(-k-1+\delta) =
  V_{a,1}(\infty) 
  - V_{a,1}^{(\delta)}(\infty) \nonumber\\
\z+\z
   (-)^a \pi{\rm ctg}(\pi\delta) \Biggl(
     W_{a-1,1}^{(-\delta)}(k)
   + W_{a}^{(-\delta)}(k)
   - \pi{\rm ctg}(\pi\delta) W_{a-1}^{(-\delta)}(k) \Biggr)
   - {2\delta\choose\delta}^{\!\!\!-1}
     \frac{\psi(\delta)+\gamma_E}{\delta^a}
\end{eqnarray}
In the same manner functions with 2 arguments are continued.
There is one shortcoming, however, in such a procedure---some of
the infinite sums are divergent, e.g. $W_a(\infty)$.
This may cause problems in some cases. One way out may be
that these functions are evaluated as hypergeometric series
with the corresponding argument, like e.g.
\begin{equation}
\label{hyperF}
W_1^{(\delta)}(\infty) = \frac{\Gamma(3+2 \delta)}{\Gamma(1+\delta) \Gamma(2+\delta)}
              ~_3F_2(1,\frac{3}{2}+\delta,1+\delta;2+\delta,2+\delta;4).
\end{equation}
Moreover,
some of them (e.g. $S_1$) cancel after applying the SW-transformation.

  As an example of this technique we transform
the series $z^n V_3/n$ into a $1/z^n$ series. Utilizing
formula (\ref{Va}) we get
\begin{eqnarray}
\label{V3overn}
\sum_{n=1}^{\infty} \frac{V_3(n-1)}{n} z^n \z=\z
     \frac{1}{24}\log^4(-z)
   + 2 \zeta_3 \log(-z)
   - 3 \zeta_4
   - 2 V_4(\infty)
   + 2 V_{3,1}(\infty)
   - 2 \tilde{V}_{3,1}(\infty)  \nonumber\\
   \z-\z \sum_{n=1}^{\infty} \frac{1}{z^n}\,\frac{1}{n}
   \Biggl( 
       \frac16\log^3(-z) 
     + \frac{1}{2n}\log^2(-z)
     + \Bigl( \frac{1}{n^2} + W_2(n) \Bigr)\log(-z) \nonumber\\
   \z+\z 2 \zeta_3 
     + 3 W_3(n)
     + 2 W_{2,1}(n)
     - 2 \tilde{W}_{2,1}(n)
     + \frac{W_2(n)}{n}
     + \frac{1}{n^3}
   \Biggr).
\end{eqnarray}
In this formula the first line is the residue at $n=0$ while
the summation is due to residua at negative $n$.

\section{Conclusion}

  For a certain class of selfenergy and vertex functions with only
one non-zero mass we have developed a new technique of obtaining
analytic results. This method uses information from the large mass
expansion of the diagrams by reproducing the obtained expansion
coefficients in terms of a certain set of `basis elements', which are
harmonic sums in many cases but often more complicated (`W- and V-sums').
However, we did not investigate systematically all $2m$-vertex
functions. In some cases we failed to find the solution.
Supposedly some of them include new exceptional basis elements.

  Given the expansion coefficients analytically, it is possible
to sum most of the series in closed form to yield a representation of
the diagrams under consideration in terms of polylogarithms. Furthermore
the analytic form of the expansion coefficients
allows by means of the Sommerfeld-Watson transformation 
a direct transition to the large $q^2$-expansion. For diagrams involving
elliptic functions (essentially those with $3m$-cuts) the basis
elements of the corresponding expansions were not found.
 
\bigskip
\noindent
{\large{\bf Acknowledgement}}

\medskip

O.V. gratefully acknowledges financial
support by the BMBF under 05 7BI92P 9, A.K. from the Volkswagen-Stiftung
and Russian Fund for Fundamental Investigations under 98-02-16923. 

%\newpage

\begin{center}
{\Large\bf Appendix}
\end{center}

\appendix

%%%%%%%%%%%%%%%%%%%%%%%%%%%%%%%%%%%%%%%%%%%%%%%%%%%%%%%%%%%%%%%
%
%  basic functions
%
%%%%%%%%%%%%%%%%%%%%%%%%%%%%%%%%%%%%%%%%%%%%%%%%%%%%%%%%%%%%%%%

\section{Basic functions}

  Here we list all basic functions we used and point
their 'transcendentality level' (TL).

\subsection{Zero-fold sums}
\begin{eqnarray}
 \z\z  \zeta_a\,,         \label{x1}            \\
 \z\z  \frac{1}{n^a}\,,   \label{x2}            \\
 \z\z  {2j\choose j}\,,   \label{x3}            \\
 \z\z  {2j\choose j}^{\!\!\!-1}\,, \label{x4}           
\end{eqnarray}
where $\zeta_a=\zeta(a)$ is the Riemann $\zeta$-function.
(\ref{x1}) and (\ref{x2}) have TL$=a$.
(\ref{x3}) and (\ref{x4}) have TL$=0$.

\subsection{One-fold sums}
\begin{eqnarray}
  S_a(n) \z=\z \sum_{j=1}^{n} \frac{1}{j^a}\,,               \\
  K_a(n) \z=\z - \sum_{j=1}^{n} \frac{(-)^j}{j^a}\,,         \\
  W_a(n) \z=\z \sum_{j=1}^{n} {2j\choose j} \frac{1}{j^a}\,, \\
  V_a(n) \z=\z \sum_{j=1}^{n} 
          {2j\choose j}^{\!\!\!-1} \frac{1}{j^a}\,. 
\end{eqnarray}
All have TL$=a$.
An exceptional sum occurs in the diagram $I_{123}$ with TL$=2$
\begin{equation}
\label{v2hat}
  \hat{V}_2(n) = \sum_{j=1}^{n} 
       {2j\choose j}^{\!\!\!-1} \frac{1}{j(n+1-j)}.
\end{equation}

\subsection{Two-fold sums}
\begin{eqnarray}
  S_{a,b}(n) \z=\z \sum_{j=1}^{n} \frac{1}{j^a} S_b(j-1) \,,      \\
  K_{a,b}(n) \z=\z - \sum_{j=1}^{n} \frac{(-)^j}{j^a} S_b(j-1)\,, \\
  W_{a,b}(n) \z=\z \sum_{j=1}^{n} 
           {2j\choose j} \frac{1}{j^a} S_b(j-1)\,,                \\
  V_{a,b}(n) \z=\z \sum_{j=1}^{n} 
           {2j\choose j}^{\!\!\!-1} \frac{1}{j^a} S_b(j-1)\,. 
\end{eqnarray}
If instead of $S_a(j-1)$ there is an insertion of $S_a(2j-1)$ we
use the tilde, e.g.
\begin{equation}
\label{v21tilde}
  \tilde{V}_{a,b}(n) = \sum_{j=1}^{n} 
      {2j\choose j}^{\!\!\!-1} \frac{1}{j^a} S_b(2j-1)\,,
\end{equation}
etc.
All have TL$=a+b$.

\subsection{Three-fold sums}
\begin{eqnarray}
  S_{a,(b+c)}(n) \z=\z \sum_{j=1}^{n} 
      \frac{1}{j^a} S_b(j-1) S_c(j-1)\,,                    \\
  W_{a,(b+c)}(n) \z=\z \sum_{j=1}^{n} 
      {2j\choose j} \frac{1}{j^a} S_b(j-1) S_c(j-1)\,,      
\end{eqnarray}
All have TL$=a+b+c$.

\subsection{Functions with two arguments}
  Functions $S$ and $V$ may occur with two arguments
\begin{eqnarray}
  S_a(n;p) \z=\z \sum_{j=1}^n \frac{p^{j-n}}{j^a}\,, \\ 
  V_a(n;p) \z=\z \sum_{j=1}^n 
   {2j\choose j}^{\!\!\!-1} \frac{p^{j-n}}{j^a}\,, 
\end{eqnarray}
All have TL$=a$.

%%%%%%%%%%%%%%%%%%%%%%%%%%%%%%%%%%%%%%%%%%%%%%%%%%%%%%%%%%%%%%%
%
%  2-point function
%
%%%%%%%%%%%%%%%%%%%%%%%%%%%%%%%%%%%%%%%%%%%%%%%%%%%%%%%%%%%%%%%
\section{The 2-point functions}
Below we present results for the diagrams shown in Fig.2
All functions have argument $n-1$ which is omitted.
Functions with two arguments have their first argument
$n-1$ which is omitted. 
$\overline{S}_1=S_1(2n-1)$ and $\hat{V}_2$ in $I_{123}$
is given by (\ref{v2hat}).
$z=q^2/m^2$.
\begin{eqnarray}
%
% 1
%
I_1 \z=\z 
  \frac{1}{q^2} 
  \sum_{n=1}^{\infty} z^n \frac{1}{n}
       \Biggl[
    \frac12 \log^2(-z) - \frac{2}{n}\log(-z)
   + \zeta_2 + 2 S_2 - 2 \frac{S_1}{n} + \frac{3}{n^2} 
       \Biggr]  \\
%
% 5
%
I_5 \z=\z 
  \frac{1}{q^2} 
  \sum_{n=1}^{\infty} z^n \frac{(-)^n}{n}
       \Biggl[
   - \log^2(-z) + \frac{2}{n}\log(-z) 
   - 2\zeta_2 + 4 K_2 - 2 \frac{1}{n^2} - 2 \frac{(-)^{n}}{n^2} 
       \Biggr]  \\
%
% 12
%
I_{12} \z=\z 
  \frac{1}{q^2} 
  \sum_{n=1}^{\infty} z^n \frac{1}{n^2}
     \Biggl[
       \frac{1}{n}
     + {2n\choose n}^{\!\!\!-1}
       \Biggl( - 2 \log(-z) - 3 W_1 + \frac{2}{n} \Biggr)
     \Biggr]  \\
% 13
I_{13} \z=\z 
  \frac{1}{q^2} 
  \sum_{n=1}^{\infty} z^n \frac{1}{n}
       \Biggl[
     2 \zeta_2 + 2 S_2 - 2\frac{S_1}{n}
       \Biggr] \\
%
% 14
%
I_{14} \z=\z 
  \frac{1}{q^2} 
  \sum_{n=1}^{\infty} z^n \frac{1}{n}
       \Biggl[
     - S_1(;2)\log(-z)  \nonumber\\
  \z\z
     + 2\zeta_2 - 6 V_2 + 2 V_2(;2) 
     + S_2(;2) - S_{1,1}(;2) + 2\frac{S_1}{n} + \frac{S_1(;2)}{n}
       \Biggr]  \\
%
% 15
%
I_{15} \z=\z 
  \frac{1}{q^2} 
  \sum_{n=1}^{\infty} z^n \frac{1}{n}
       \Biggl[
     - \frac{1}{n}\log(-z) - \zeta_2
     + 3 V_2 + \frac{2}{n^2} + 4 {2n\choose n}^{\!\!\!-1} \frac{1}{n^2}
       \Biggr] \\
%
% 123
%
I_{123} \z=\z 
  \frac{1}{q^2} 
  \sum_{n=1}^{\infty} z^n \frac{1}{n}
       \Biggl[
         \Biggl(
      \zeta_2 - 3 V_2 + 3 \hat{V}_2
          \Biggr)
    - {2n\choose n}^{\!\!\!-1} 3 \frac{W_1}{n}
       \Biggr] \\
%
% 125
%
I_{125} \z=\z 
  \frac{1}{q^2} 
  \sum_{n=1}^{\infty} z^n \frac{1}{n^2} {2n\choose n}^{\!\!\!-1} 
       \Biggl[
    - \log(-z) + \frac{3}{n}
       \Biggr]  \\
%
% 1234
%
I_{1234} \z=\z 
  \frac{1}{q^2} 
  \sum_{n=1}^{\infty} z^n \frac{1}{n}
     \Biggl[
        V_2(;4)
      + {2n\choose n}^{\!\!\!-1} \frac{1}{n}
       \Biggl( -8 S_1 + 4 \overline{S}_1 - \frac{2}{n} \Biggr)
     \Biggr]
\end{eqnarray}

  First seven coefficients of the expansion for 
$I_1,\,I_5,\,I_{12},\,I_{14},\,I_{15}$ and $I_{125}$ were 
compared with \cite{propagatorbis}.

%%%%%%%%%%%%%%%%%%%%%%%%%%%%%%%%%%%%%%%%%%%%%%%%%%%%%%%%%%%%%%%
%
%  2-point integrals
%
%%%%%%%%%%%%%%%%%%%%%%%%%%%%%%%%%%%%%%%%%%%%%%%%%%%%%%%%%%%%%%%

\section{Integral representations for the 2-point functions}

Here we give the representations for the 2-point functions.
Following \cite{Devoto} we introduce the notation
\begin{equation}
\label{Sqp}
  S_{a+1,b}(z) = \frac{(-1)^{a+b}}{a!\,b!}
    \int_0^1 \frac{\log^a(t)\log^b(1-zt)}{t}\,dt.
\end{equation}

  All polylogarithmic functions are particular
cases of $S$ functions, namely
\begin{equation}
\label{Li}
  {\rm Li}_a(z) = S_{a-1,1}(z)\,.
\end{equation}

  We introduce the following new variable ($z=q^2/m^2$)
\begin{equation}
y=\frac{1-\sqrt{z/(z-4)}}{1+\sqrt{z/(z-4)}}\,.
\end{equation}

  Then
\begin{eqnarray}
q^2 \cdot I_1 \z=\z 
     - \frac{1}{2}\log^2(-z)\log(1-z) 
     - 2\log(-z){\rm Li}_2(z)
     + 3{\rm Li}_3(z) -6 S_{1,2}(z)  \nonumber \\
   \z-\z \log(1-z) \biggl(  \zeta_2 + 2{\rm Li}_2(z) \biggr)\,, \\
\z\z\nonumber\\
q^2 \cdot I_5 \z=\z 
      2 \zeta_2 \log(1+z) + 2\log(-z){\rm Li}_2(-z) + \log^2(-z)\log(1+z)
        + 4\log(1+z){\rm Li}_2(z) 
            \nonumber \\
      \z-\z 2{\rm Li}_3(-z) 
           - 2{\rm Li}_3(z) 
           + 2 S_{1,2}(z^2) - 4S_{1,2}(z) - 4 S_{1,2}(-z)\,, \\
\z\z\nonumber\\
q^2 \cdot I_{12} \z=\z
     {\rm Li}_3(z) - 6 \zeta_3 - \zeta_2\log y
     - \frac16\log^3 y - 4\log y\,{\rm Li}_2(y)  \nonumber\\
     \z+\z 4{\rm Li}_3(y) - 3{\rm Li}_3(-y) + \frac13 {\rm Li}_3(-y^3)\,,\\
\z\z\nonumber\\
q^2 \cdot I_{13} \z=\z 
      - 6 S_{1,2}(z)
      - 2 \log(1-z) \biggl(  \zeta_2 + {\rm Li}_2(z) \biggr)\,,  \\
\z\z\nonumber\\
q^2 \cdot I_{14} \z=\z  
     \log(2-z) \biggl( \log^2(1-z) -2 \log(-z)\log(1-z) -2{\rm Li}_2(z)
       \biggr)\nonumber \\
  \z-\z \frac{2}{3} \log^3(1-z) + \log(-z)\log^2(1-z)
     - 2 \zeta_2 \log(1-z) \nonumber\\
  \z-\z  S_{1,2}\bigr(1/(1-z)^2\bigl)
     + 2 S_{1,2}\bigr(1/(1-z)\bigl) + 2 S_{1,2}\bigr(-1/(1-z)\bigl)  \nonumber\\
  \z+\z \frac13 \log^3 y 
   +\log^2 y\,\biggl( 2 \log(1+y^2) -3 \log(1-y +y^2) \biggr) 
                   \nonumber \\
\z-\z 6 \zeta_3 - {\rm Li}_3(-y^2)
     +\frac23 {\rm Li}_3(-y^3) - 6 {\rm Li}_3(-y) 
         \nonumber \\ 
\z+\z 2\log y\, \biggl( {\rm Li}_2(-y^2) -{\rm Li}_2(-y^3) 
              +3 {\rm Li}_2(-y)    \biggr)\,, \\
\z\z\nonumber\\
q^2 \cdot I_{15} \z=\z 
   2{\rm Li}_3(z) - \log(-z)\,{\rm Li}_2(z) 
              + \zeta_2 \log(1-z) \nonumber \\
  \z+\z \frac16 \log^3 y 
   - \frac12\log^2 y\,\biggl( 8 \log(1-y) -3 \log(1-y +y^2) \biggr) 
                   \nonumber \\
\z-\z 6 \zeta_3 - \frac13 {\rm Li}_3(-y^3) + 3 {\rm Li}_3(-y) 
            + 8 {\rm Li}_3(y) \nonumber \\ 
\z+\z \log y\, \biggl( {\rm Li}_2(-y^3) -3 {\rm Li}_2(-y) 
              -8 {\rm Li}_2(y)   \biggr)\,, \\
\z\z\nonumber\\
q^2 \cdot I_{123} \z=\z 
    - \zeta_2 \biggl(  \log(1-z) +  \log y \biggr)
    - 6 \zeta_3  -\frac32 \log(1-y +y^2)\, \log^2 y \nonumber \\
    \z+\z {\rm Li}_3(-y^3) - 9 {\rm Li}_3(-y)
    - 2 \log y\, \biggl( {\rm Li}_2(-y^3) - 3{\rm Li}_2(-y) \biggr)  
      \,,\\
\z\z\nonumber\\
q^2 \cdot I_{125} \z=\z  
      -2 \log^2 y\, \log(1-y)
      - 6 \zeta_3 + 6 {\rm Li}_3(y) - 6\log y\, {\rm Li}_2(y)\,, \\
\z\z\nonumber\\
q^2 \cdot I_{1234} \z=\z 
    - 6 \zeta_3  - 12 {\rm Li}_3(y) - 24 {\rm Li}_3(-y)
                      \nonumber \\
    \z+\z 8 \log y\, \biggl( {\rm Li}_2(y) + 2{\rm Li}_2(-y) \biggr)  
      + 2 \log^2 y\, \biggl( \log(1-y) + 2 \log(1+y) \biggr)\,. 
\end{eqnarray}

   Different analytic representations for $I_1,\,I_{13},\,I_{125}$
and $I_{1234}$ were obtained in \cite{BroadhurstZP47}.
In \cite{BFT} the same four integrals were calculated for
arbitrary dimension $d$ in terms of hypergeometric functions. 
$I_5$ and $I_{12}$ were calculated also in \cite{SE2bis}.
The numerical check shows agreement for these integrals.

%%%%%%%%%%%%%%%%%%%%%%%%%%%%%%%%%%%%%%%%%%%%%%%%%%%%%%%%%%%%%%%
%
%  3-point function
%
%%%%%%%%%%%%%%%%%%%%%%%%%%%%%%%%%%%%%%%%%%%%%%%%%%%%%%%%%%%%%%%
\section{3-point functions}

Below we present results for the diagrams shown in Fig.3
All functions have argument $n-1$ which is omitted.
Overlined functions have argument $2n-1$ which is omitted 
($\overline{S}_1=S_1(2n-1)$ etc.).
$\tilde{V}_{2,1}$ in $P_{56}$ is given by (\ref{v21tilde}) 
and $z=q^2/m^2$.
\begin{eqnarray}
%
%  case P1
%
P_1 \z=\z  
   \frac{1}{(q^2)^2}
    \sum_{n=1}^{\infty} z^n \frac{1}{n}  \nonumber\\
  \z\z  \Biggl\{
   - \frac{1}{2\varepsilon^3}
   - \frac{1}{\varepsilon^2} S_1
   +\frac{1}{\varepsilon} \Bigl[
          - \frac12 \zeta_2
          + \frac52 S_2
          - \frac12 S_1^2
          + \frac{1}{n^2}
          - \frac{1}{n} \log(-z)
          + \frac12 \log^2(-z)
                          \Bigr]  \nonumber\\
  \z\z   
       - \frac83 \zeta_3 
       - \zeta_2 S_1 - \frac{\zeta_2}{n}
       + \frac83 S_3
       + \frac92 S_1 S_2
       + \frac56 S_1^3
       + 4 \frac{S_2}{n}
       + 2 \frac{S_1}{n^2}
       + \frac{3}{n^3} \nonumber\\
  \z\z
       + \Bigl(
           \zeta_2
         - 4 S_2
         - 2 \frac{S_1}{n}
         - \frac{3}{n^2}
         \Bigr) \log(-z)
       + \Bigl(
           S_1
         + \frac32 \frac{1}{n}
         \Bigr) \log^2(-z)
       - \frac12 \log^3(-z)
      \Biggr\}  \\
%
%  case P3
%
P_3 \z=\z  
   \frac{1}{(q^2)^2}
    \sum_{n=1}^{\infty} z^{n}  \nonumber\\
  \z\z  \Biggl\{
    \frac{1}{\varepsilon^2} \Bigl[ \frac12\zeta_2 - \frac12 S_2 \Bigr]
   + \frac{1}{\varepsilon} \Bigl[
             \frac12\zeta_3
           + \zeta_2 S_1
           - \frac52 S_3
           - S_{1,2}
           + \Bigl(
              - \zeta_2
              + S_2
             \Bigr) \log(-z)
                             \Bigr] \nonumber\\
 \z\z
      + \zeta_4
      + \zeta_3 S_1
      + \frac72 \zeta_2 S_2
      + \zeta_2 S_1^2
      + \frac32 S_4
      + 3 S_{1,3}
      - 2 S_{1,(1+2)}
      - 8 S_1 S_3
      - 2 S_1 S_{1,2}
      - 2 S_2^2
      + 2 S_1^2 S_2
       \nonumber\\
 \z\z
      + \Bigl(
         - \zeta_3
         - 2 \zeta_2 S_1
         + S_3
         + 2 S_1 S_2
        \Bigr) \log(-z) 
      + \Bigl(
           \zeta_2
         - \frac12 S_2
        \Bigr) \log^2(-z)
      \Biggr\}  \\
%
%  case P5
%
P_5 \z=\z  
   \frac{1}{(q^2)^2}
    \sum_{n=1}^{\infty} z^n \frac{(-)^n}{n}  \nonumber\\
  \z\z  \Biggl\{
      - 6 \zeta_3
      + 2 \zeta_2 S_1
      + 6 S_3
      - 2 S_1 S_2
      + 4 \frac{S_2}{n}
      - \frac{S_1^2}{n}
      + 2 \frac{S_1}{n^2} \nonumber\\
 \z\z
      + \Bigl(
        - 4 S_2
        + S_1^2
        - 2 \frac{S_1}{n}
        \Bigr) \log(-z)
      + S_1 \log^2(-z)
      \Biggr\}  \\
%
%  case P6
%
P_6 \z=\z  
   \frac{1}{(q^2)^2}
    \sum_{n=1}^{\infty} z^n \frac{(-)^n}{n}  \nonumber\\
  \z\z  \Biggl\{
    \frac{1}{\varepsilon^2} \Bigl[ - \frac{1}{n} + \log(-z) \Bigr]
                      \nonumber\\
  \z\z
   + \frac{1}{\varepsilon} \Bigl[
             \zeta_2
           - 3 S_2
           + 4 K_2
           - 3 \frac{S_1}{n}
           - \frac{3}{n^2}
           + \Bigl(
                3 S_1
              + \frac{3}{n}
             \Bigr) \log(-z)
           - \frac32 \log^2(-z)
                             \Bigr] \nonumber\\
 \z\z
      + 2 \zeta_3   
      + 7 \zeta_2 S_1 
      + 2 \frac{\zeta_2}{n} 
      - 2 S_3
      - 9 S_1 S_2
      + 2 K_3
      + 12 K_{2,1}
      + 4 S_1 K_2       
      - \frac72 \frac{S_2}{n} \nonumber\\
 \z\z       
      - \frac92 \frac{S_1^2}{n}       
      - 5 \frac{S_1}{n^2}
      - \frac{7}{n^3}       
      + \Bigl(
         - 2 \zeta_2 
         + \frac72 S_2
         + \frac92 S_1^2
         + 5 \frac{S_1}{n}
         + \frac{7}{n^2}
        \Bigr) \log(-z)   \nonumber\\
 \z\z
      + \Bigl(
         - \frac52 S_1
         - \frac72 \frac{1}{n}
        \Bigr) \log^2(-z)
      + \frac76 \log^3(-z)
      \Biggr\}  \\
%
%  case P13
%
P_{13} \z=\z  
   \frac{1}{(q^2)^2}
    \sum_{n=1}^{\infty} z^n  \nonumber\\
  \z\z  \Biggl\{
  - \frac{1}{\varepsilon^2}\,\frac12 S_2
  + \frac{1}{\varepsilon} \Bigl[
           - \frac12 S_3
           - 2 S_{1,2}
            \Bigr] \nonumber\\
 \z\z
      + \frac12 \zeta_2 S_2
      - \frac52 S_4
      - S_{1,3}
      + 3 S_{1,(1+2)}
      - S_1 S_3
      - 8 S_1 S_{1,2}
      + \frac12 S_2^2
      + \frac52 S_1^2 S_2
      \Biggr\}  \\
%
%  case P12
% 
P_{12} \z=\z  
   \frac{1}{(q^2)^2}
    \sum_{n=1}^{\infty} z^n {2n\choose n}^{\!\!\!-1}\frac{1}{n^2}\nonumber\\
  \z\z  \Biggl\{
     \frac{2}{\varepsilon^2}
    +\frac{1}{\varepsilon} \Bigl[
      - 6 W_1
      + 2 S_1
      + \frac{2}{n}
      - 2 \log(-z)
           \Bigr] 
      - 6 W_2
      - 18 W_{1,1}
      - 13 S_2     \nonumber\\
 \z\z
      + S_1^2   
      - 6 S_1 W_1
      + 2 \frac{S_1}{n}
      + \frac{2}{n^2}
      + \Bigl(
         - 2 S_1
         - \frac{2}{n}
        \Bigr) \log(-z)
      +  \log^2(-z)
      \Biggr\} \\
%
%  case P56
%
P_{56} \z=\z  
   \frac{1}{(q^2)^2}
    \sum_{n=1}^{\infty} z^n \frac{(-)^n}{n}  \nonumber\\
  \z\z  \Biggl\{
      - 6 \zeta_2 S_1 
      + 6 V_3
      - 2 S_3
      - 10 K_3
      - 12 V_{2,1}
      + 12 \tilde{V}_{2,1}
      - 12 K_{2,1}
      + 12 S_1 K_2       
      + 6 \frac{V_2}{n}   \nonumber\\
 \z\z    
      - 3\frac{S_2}{n}
      - 4\frac{S_1}{n^2}
      +\Bigl(
        - 6 V_2
        + 3 S_2
        + 4 \frac{S_1}{n}
        \Bigr) \log(-z)
      - 2 S_1 \log^2(-z)
      \Biggr\}  \\
%
%  case P126
%
P_{126} \z=\z  
   \frac{1}{(q^2)^2}
    \sum_{n=1}^{\infty} z^n {2n\choose n}^{\!\!\!-1}\frac{1}{n^2}\nonumber\\
  \z\z  \Biggl\{
      \frac{1}{\varepsilon^2}
    + \frac{1}{\varepsilon} \Bigl[ 
           - S_1
           - \log(-z)
             \Bigr] 
      - \frac32 S_2
      - \frac{15}{2} S_1^2
      + 4 S_1 \overline{S}_1  \nonumber\\
  \z\z  + 2 \frac{S_1}{n}
      - S_1 \log(-z)
      + \frac12 \log^2(-z)
      \Biggr\}  \\
%
%  case P3456
%
P_{3456} \z=\z 
  \frac{1}{(q^2)^2} 
  \sum_{n=1}^{\infty} z^n {2n\choose n}^{\!\!\!-1}\frac{1}{n} \nonumber\\
  \z\z  \Biggl\{
        7 S_3
      - 4 S_{1,2}
      - 8 K_3
      + 8 S_1 K_2
      + 8 \overline{S}_3
      + 8 \overline{S}_{1,2}
      + 8 \overline{K}_{2,1}
      + 4 \overline{S}_1 S_2
      - 8 \overline{S}_1 K_2
      - 8 \overline{S}_1 \overline{S}_2 \nonumber\\
 \z\z
      + \Bigl(
         - 2 S_2
         + 4 K_2
        \Bigr) \log(-z)
      \Biggr\} \\
%
%  case N1
%
N_1 \z=\z 
   \frac{1}{(q^2)^2}
    \sum_{n=1}^{\infty} z^n  \nonumber\\
  \z\z  \Biggl\{
      \frac{1}{4\varepsilon^4} 
    + \frac{1}{\varepsilon^3} \Biggl[
              S_1
            - \frac12 \log(-z)
                              \Biggr]
    + \frac{1}{\varepsilon^2} \Biggl[
            - \frac14 \zeta_2
            +  3 S_2
            + 2 S_1^2
            - 2 S_1 \log(-z)
            + \frac12 \log^2(-z)
                              \Biggr]  \nonumber\\
 \z\z
    + \frac{1}{\varepsilon}   \Biggl[
          - \frac83 \zeta_3
          - \zeta_2 S_1     
          + \frac{10}{3} S_3
          + 2 S_{1,2}
          + 10 S_1 S_2
          + \frac83 S_1^3  \nonumber\\
 \z\z
          + \Bigl(
               \frac12 \zeta_2
             - 4 S_2
             - 4 S_1^2
            \Bigr) \log(-z)
          + 2 S_1 \log^2(-z)
          - \frac13 \log^3(-z)
                             \Biggr]  \nonumber\\
 \z\z
      - \frac{57}{16} \zeta_4  
      - \frac{32}{3} \zeta_3 S_1
      + \zeta_2 S_2 
      - 2 \zeta_2 S_1^2
      + 4 S_4
      - 4 S_{1,3}
      - 2 S_{1,(1+2)}
      + 7 S_2^2
      + \frac{52}{3} S_1 S_3    \nonumber\\
  \z\z
      + 8 S_1 S_{1,2}
      + 17 S_1^2 S_2
      + \frac83 S_1^4
      + \Bigl(
           \frac{16}{3} \zeta_3
         + 2 \zeta_2 S_1
         - \frac{20}{3} S_3
         - 16 S_1 S_2
         - \frac{16}{3} S_1^3
        \Bigr) \log(-z)  \nonumber\\       
 \z\z
      + \Bigl(
         - \frac12 \zeta_2
         + 4 S_2
         + 4 S_1^2
        \Bigr) \log^2(-z)      
      - \frac43 S_1 \log^3(-z)
      + \frac16 \log^4(-z)
     \Biggr\} \\
%
%  case N3
% 
N_3 \z=\z  
   \frac{1}{(q^2)^2}
    \sum_{n=1}^{\infty} z^n (-)^n  \nonumber\\
  \z\z  \Biggl\{
         \frac{1}{\varepsilon^2} \Bigl[
                 - \frac12 \zeta_2
                 + K_2 
                                \Bigr]
      +\frac{1}{\varepsilon}  \Bigl[
               - \frac12 \zeta_3 
               - 2 \zeta_2 S_1
               + 2 S_3
               - 2 K_3
               + 4 S_1 K_2
               + \Bigl( \zeta_2 - S_2 \Bigr) \log(-z)
                              \Bigr]   \nonumber\\
  \z\z
        - \zeta_4
        - 2 \zeta_3 S_1
        - 7 \zeta_2 S_2 
        - 4 \zeta_2 S_1^2 
        + 7 \zeta_2 K_2 \nonumber\\
   \z\z
        - \frac72 S_4
        + \frac72 S_2^2
        + 6 S_1 S_3
        + 2 S_{1,3}
        + 8 K_4
        - 8 S_1 K_3
        + 8 S_1^2 K_2   \nonumber\\
  \z\z
        + \Bigl( 
                      \zeta_3 
                    + 4 \zeta_2 S_1
                    - S_{1,2}
                    - 3 S_1 S_2
                    - 4 K_3 
          \Bigr) \log(-z)
        + \Bigl( 
                    - \zeta_2
                    + \frac12 S_2
                    + K_2 
          \bigr) \log^2(-z)
     \Biggr\}  \\
%
%  case N13
%
N_{13} \z=\z  
   \frac{1}{(q^2)^2}
    \sum_{n=1}^{\infty} z^n \frac{1}{n}  \nonumber\\
  \z\z  \Biggl\{
    - \frac{1}{\varepsilon^2}\,S_1
    + \frac{1}{\varepsilon}  \Bigl[
        - \frac12 S_2
        - \frac72 S_1^2
        + 2 \frac{S_1}{n}
                            \Bigr] \nonumber\\
  \z\z
     - 5 \zeta_2 S_1
     + \frac{14}{3} S_3
     + 2 S_{1,2}
     - \frac12 S_1 S_2
     - \frac{37}{6} S_1^3
     + \frac{S_2}{n}
     + 7 \frac{S_1^2}{n}
     - 4 \frac{S_1}{n^2}
      \Biggr\} \\
%
%  case N12
%
N_{12} \z=\z  
   \frac{1}{(q^2)^2}
    \sum_{n=1}^{\infty} z^n {2n\choose n}^{\!\!\!-1}\frac{1}{n} \nonumber\\
  \z\z  \Biggl\{
    - \frac{6}{\varepsilon^2} W_1 
    +\frac{1}{\varepsilon} \Bigl[
          - 4 W_2
          - 12 W_{1,1}
          - 12 S_1 W_1
          - 16 S_2
               \Bigr]  \nonumber\\
 \z\z
      - 18 \zeta_2 W_1
      - 4 W_3
      - 8 S_3
      + 12 W_{1,2}
      - 8 W_{2,1}
      - 12 W_{1,(1+1)}  \nonumber\\
 \z\z
      - 16 S_{1,2} 
      - 8 S_1 W_2
      - 24 S_1 W_{1,1}
      - 16 S_1 S_2
      - 12 S_1^2 W_1
      \Biggr\}
\end{eqnarray}

%%%%%%%%%%%%%%%%%%%%%%%%%%%%%%%%%%%%%%%%%%%%%%%%%%%%%%%%%%%%%%%%%%%%%%

\section{Useful sums}

  In Appendix B we have presented our results for two-point
functions in terms of the basis elements given in appendix A.
Summing the series, we were able to obtain closed expressions
in terms of polylogarithms. 
In appendix D
we also gave the series representation for the vertex functions
under consideration. The summation and representation in terms
of integrals of polylogarithms is done for separate structures,
occuring repeatedly in these series.

   In the case when only harmonic
sums enter the expression the formulae given in \cite{Devoto}
are very useful. For convenience we introduce the notation

\begin{equation}
  \Li_1(z) = -\log(1-z).
\end{equation}

\subsection{Sums involving $S$ and $K$}

The following general relations hold true
\begin{eqnarray}
\z\z\sum_{n=1}^{\infty} \frac{f(n-1)}{n^a}\,z^n = 
\int^z_0 \frac{dz'}{z'} 
\sum_{n=1}^{\infty} \frac{f(n-1)}{n^{a-1}}\,(z')^n\,,  \\
\z\z\sum_{n=1}^{\infty} f(n-1)\,z^n = \frac{z}{1-z}
\sum_{n=1}^{\infty} \biggl[ f(n) - f(n-1) \biggl]\,z^n,  
\end{eqnarray}
where $f(n-1)$ is any function. 

  As a particular result we have
\begin{eqnarray}
\z\z\sum_{n=1}^{\infty} \frac{1}{n^a}\,z^n = \Li_a(z)\,, \\
\z\z\sum_{n=1}^{\infty} S_a(n-1)\,z^n = \frac{z}{1-z}\Li_a(z)\,, \\
\z\z\sum_{n=1}^{\infty} \frac{S_1(n-1)}{n^a}\,z^n = S_{a-1,2}(z)\,, \\
\z\z\sum_{n=1}^{\infty} K_a(n-1)\,z^n = \frac{z}{1-z}\Li_a(-z)\,. 
\end{eqnarray}
 
 Only $S$-functions (for brevity: on the l.h.s. $S_a=S_a(n-1)$ etc. and 
on the r.h.s. $\Li_a = \Li_a(z)$ etc.) 
\begin{eqnarray}
%
%  with 1/n
%
\z\z\sum_{n=1}^{\infty} \frac{S_2}{n}\,z^n = 
         \Li_1\Li_2 - 2S_{1,2},  \\
\z\z\sum_{n=1}^{\infty} \frac{S_1^2}{n}\,z^n = 
         \Li_1\Li_2 + \frac13\Li_1^3 - 2S_{1,2},  \\
\z\z\sum_{n=1}^{\infty} \frac{S_3}{n}\,z^n = 
         \Li_1\Li_3 - \frac12\Li_2^2,  \\
\z\z\sum_{n=1}^{\infty} \frac{S_1S_2}{n}\,z^n = 
         \frac12\Li_1^2\Li_2 + \Li_1\Li_3 
       - \frac12\Li_2^2 - \Li_1 S_{1,2},  \\
\z\z\sum_{n=1}^{\infty} \frac{S_1^3}{n}\,z^n = 
        \frac14\Li_1^4 + \Li_1\Li_3 -  \frac12\Li_2^2
       + \frac32\Li_1^2\Li_2 - 3\Li_1 S_{1,2}, \\
\z\z\sum_{n=1}^{\infty} \frac{S_{1,2}}{n}\,z^n = 
         \frac12\Li_1^2\Li_2 - 2\Li_1 S_{1,2} + 3S_{1,3}, \\
%
%  with 1/n^2
%
\z\z\sum_{n=1}^{\infty} \frac{S_2}{n^2}\,z^n = 
         \frac12\Li_2^2 - 2 S_{2,2},  \\
\z\z\sum_{n=1}^{\infty} \frac{S_1^2}{n^2}\,z^n = 
         \frac12\Li_2^2 + 2S_{1,3} - 2S_{2,2},  \\
%
%  without 1/n
%
\z\z\sum_{n=1}^{\infty} S_1^2\,z^n = 
        \frac{z}{1-z} \Biggl( \Li_1^2 + \Li_2 \Biggr), \\
\z\z\sum_{n=1}^{\infty} S_1 S_2\,z^n = 
        \frac{z}{1-z} \Biggl( \Li_3 + \Li_1 \Li_2 - S_{1,2} \Biggr), \\
\z\z\sum_{n=1}^{\infty} S_{1,2}\,z^n = 
        \frac{z}{1-z} \Biggl( \Li_1 \Li_2 - 2S_{1,2} \Biggr), \\
\z\z\sum_{n=1}^{\infty} S_1^3\,z^n = 
        \frac{z}{1-z} \Biggl( \Li_3 + 3\Li_1 \Li_2
                       + \Li_1^3 - 3S_{1,2} \Biggr), \\
\z\z\sum_{n=1}^{\infty} S_1 S_3\,z^n = 
        \frac{z}{1-z} \Biggl( \Li_4 + \Li_1 \Li_3
                      - \frac12 \Li_2^2 + S_{2,2} \Biggr), \\
\z\z\sum_{n=1}^{\infty} S_2^2\,z^n = 
        \frac{z}{1-z} \Biggl( \Li_2^2 + \Li_4 - 4 S_{2,2}  \Biggr), \\
\z\z\sum_{n=1}^{\infty} S_1^2 S_2\,z^n = 
        \frac{z}{1-z} \Biggl( \Li_1^2 \Li_2 + \Li_4 + 2\Li_1 \Li_3
        -2 \Li_1 S_{1,2} +2 S_{1,3} -2 S_{2,2}  \Biggr), \\
\z\z\sum_{n=1}^{\infty} S_1^4\,z^n = 
        \frac{z}{1-z} \Biggl( \Li_2^2 + \Li_4 +4 \Li_1 \Li_3
+6 \Li_1^2 \Li_2 + \Li_1^4 -12 \Li_1 S_{1,2} +12 S_{1,3} -8 S_{2,2}
                        \Biggr), \\
\z\z\sum_{n=1}^{\infty} S_{1,3}\,z^n = 
        \frac{z}{1-z} \Biggl( \Li_1\Li_3 - \frac12\Li_2^2 
                          \Biggr), \\
\z\z\sum_{n=1}^{\infty} S_{1,(1+2)}\,z^n = 
        \frac{z}{1-z} \Biggl( \Li_1\Li_3 + \frac12\Li_1^2\Li_2
                      -\frac12\Li_2^2 - \Li_1 S_{1,2}  \Biggr). 
\end{eqnarray}
%file chtable.form to check.

  $S$ and $K$
\begin{eqnarray}
%
%  with 1/n
%
\z\z\sum_{n=1}^{\infty} \frac{K_2}{n}\,z^n = 
                        \phi(z),   \\
\z\z\sum_{n=1}^{\infty} \frac{K_{2,1}}{n}\,z^n = 
          -\int_0^z\frac{S_{1,2}(-t)}{1-t}\,dt,  \\
\z\z\sum_{n=1}^{\infty} \frac{K_3}{n}\,z^n = 
          -\int_0^z\frac{\Li_3(-t)}{1-t}\,dt,  \\
\z\z\sum_{n=1}^{\infty} \frac{S_1 K_2}{n}\,z^n = 
          -\log(1-z)\phi(z)
          -\int_0^z
           \frac{\log(1-t)\Li_2(-t)+\Li_3(-t)+S_{1,2}(-t)}{1-t}\,dt,  \\
%
%  with 1/n^2
%
\z\z\sum_{n=1}^{\infty} \frac{K_2}{n^2}\,z^n = 
          \phi(z)\log z + \int_0^z \frac{\log t\,\Li_2(-t)}{1-t}\,dt, \\
%
%  without 1/n
%
\z\z\sum_{n=1}^{\infty} K_{2,1}\,z^n = 
        \frac{z}{1-z} \Biggl( -S_{1,2}(-z) \Biggr), \\
\z\z\sum_{n=1}^{\infty} S_1 K_2\,z^n = 
        \frac{z}{1-z} \Biggl( \phi(z)-S_{1,2}(-z)-\Li_3(-z) \Biggr), \\
\z\z\sum_{n=1}^{\infty} S_1 K_3\,z^n = 
        \frac{z}{1-z} \Biggl( -\Li_4(-z) - S_{2,2}(-z)
                 -\int_0^z\frac{\Li_3(-t)}{1-t}\,dt \Biggr), \\
\z\z\sum_{n=1}^{\infty} S_2 K_2\,z^n = 
        \frac{z}{1-z} \Biggl( -\Li_4(-z) - \frac12\Li_2^2(-z) 
          + 2 S_{2,2}(-z) + \phi(z)\log z \nonumber \\
\z\z +
\int_0^z\frac{\log t \Li_2(-t)}{1-t}\,dt \Biggr), \\
\z\z\sum_{n=1}^{\infty} K_{3,1}\,z^n = 
        \frac{z}{1-z} \Biggl( -S_{2,2}(-z) \Biggr), \\
\z\z\sum_{n=1}^{\infty} S_1^2 K_2\,z^n = 
        \frac{z}{1-z} \Biggl( 
                  -\Li_4(-z) - \frac12\Li_2^2(-z) - 2S_{1,3}(-z) 
     +\phi(z)\Bigl(\log z + 2 \log (1-z) \Bigr) \nonumber \\
   \z\z ~~~- \int_0^z \frac{\bigl(2\log (1-t)-\log t \bigr) \Li_2(-t)
  + 2 \Li_3(-t) + 2 S_{1,3}(-t)}{1-t}\,dt  \Biggr)\,, 
\end{eqnarray}
where
\begin{equation}
\phi(z) = \frac12 S_{1,2}(z^2)-S_{1,2}(z)-S_{1,2}(-z)+\log(1-z){\rm Li}_2(-z).
\end{equation}

\subsection{Sums involving $V$ and $W$}
The factor ${2n\choose n}^{\!\!\!-1} $ can be written in the following
useful form

\begin{eqnarray}
{2n\choose n}^{\!\!\!-1} = \frac{n}{2} \int^1_0 ds s^{n-1} (1-s)^{n-1}
\nonumber \end{eqnarray}

  Again we make use of some general formulae.\\

For any function $f$ we have (hereafter $p=s(1-s)$)
\begin{eqnarray}
\z\z \sum_{n=1}^{\infty} {2n\choose n}^{\!\!\!-1} \frac{f(n-1)}{n}\,z^n = 
\int^1_0 \frac{ds}{s} 
\sum_{n=1}^{\infty} f(n-1)\,(zp)^n\,,  \nonumber\\
\z\z \sum_{n=1}^{\infty} {2n\choose n}^{\!\!\!-1} \frac{f(n-1)}{n}
\Bigl[S_1 -\overline S_1 \Bigr]\,z^n = \frac{1}{2} 
\int^1_0 \frac{ds}{s} \log (p)
\sum_{n=1}^{\infty} f(n-1)\,(zp)^n\,, \nonumber\\
\z\z\sum_{n=1}^{\infty}  {2n\choose n}^{\!\!\!-1}\frac{f(2n-1)}{n}
\,z^n = \frac{1}{2} 
\int^1_0 \frac{ds}{s} 
\sum_{n=1}^{\infty} f(n-1)\Bigl[ \,(\sqrt{zp})^{n} + \,(-\sqrt{zp})^{n} \Bigr]\,.
\end{eqnarray}

  When the function $V_{a,b}(n-1)$ contribute we have
\begin{eqnarray}
\z\z\sum_{n=1}^{\infty} \frac{V_2}{n}\,z^n = 
\int^1_0 \frac{ds}{s} I_2(z,p)  \nonumber \\
\z\z\sum_{n=1}^{\infty} \frac{1}{n}
\biggl[ V_2 + \hat V_2 \biggr]\,z^n = 
\int^1_0 \frac{ds}{s} \log (1-z) \log (1-pz)  \nonumber \\
\z\z\sum_{n=1}^{\infty} \frac{V_{2,1}}{n}\,z^n = 
  \int^1_0 \frac{ds}{s} \biggl[ -I_2(z,p)\log (1-p) + I_{2,1}(z,p)
\biggr] \nonumber \\
\z\z\sum_{n=1}^{\infty} \frac{1}{n}
\biggl[ V_{2,1} - \tilde V_{2,1} \biggr]\,z^n = 
 \frac{1}{2} \int^1_0 \frac{ds}{s} \log (p)  I_2(z,p)  \nonumber \\
\z\z\sum_{n=1}^{\infty} \frac{V_{3}}{n}\,z^n = 
- \int^1_0 \frac{ds}{s} \biggl[ \log (1-z) \Li_2(zp)+ S_{1,2}(z)+ S_{1,2}(zp)
 + I_{3}(z,p)
\biggr] \nonumber \\
\z\z\sum_{n=1}^{\infty} \frac{V_{2}}{n^2}\,z^n = 
\int^1_0 \frac{ds}{s} \biggl[ I_2(z,p)\log (z) +
\log (1-p) \Bigl( \Li_2(1-z) - \zeta_2 \Bigr)\nonumber \\
\z\z~~~ - \frac{1}{2}
\log^2 (1-p)\log (1-z) + I_{2,1}(z,p)
+ I_{1,2}(z,p) \biggr], \nonumber
\end{eqnarray}
where 
\begin{eqnarray}
\z\z I_2(z,p)=\Li_2(\frac{p(1-z)}{1-zp}) - \Li_2(p) - 
\log (\frac{1-zp}{1-z})\log (1-p) + \frac{1}{2}\log^2(1-zp) \nonumber \\
\z\z I_3(z,p)= \Li_3(p) -\zeta_3- \Li_3(\frac{p(1-z)}{1-zp}) +
\Li_3(\frac{1-z}{1-zp}) \nonumber \\
\z\z~~~ + 
\log (\frac{1-z}{1-zp}) \biggl[
\Li_2(\frac{p(1-z)}{1-zp}) - \Li_2(\frac{1-z}{1-zp}) \biggr] -
 \frac{1}{2}\log (z) \log^2(\frac{1-z}{1-zp}) \nonumber \\
\z\z I_{2,1}(z,p)=S_{1,2}(p) - S_{1,2}(\frac{p(1-z)}{1-zp}) +
 \frac{1}{2}\log^2 (1-p) \log (\frac{1-z}{1-zp})
+  \frac{1}{2}\log (1-p) \log^2(1-zp) \nonumber \\
\z\z~~~- \frac{1}{6} \log^3(1-zp) \nonumber \\
\z\z I_{1,2}(z,p)= S_{1,2}(\frac{1-z}{1-zp}) - \Li_3(pz)
+ S_{1,2}(pz) - S_{1,2}(1-z) + \log (z(1-p))\Li_2(zp)\nonumber \\
\z\z~~~ +
 \frac{1}{2}\log (1-zp) \log^2(z(1-p))
-  \frac{1}{2}\log (z(1-p)) \log^2(1-zp) + \frac{1}{6}\log^3(1-zp)\nonumber 
\end{eqnarray}

  When the functions $W_{a,b}(n-1)$ contribute we have
\begin{eqnarray}
\z\z 3\sum_{n=1}^{\infty}{2n\choose n}^{\!\!\!-1}
 \frac{1}{n} W_1 \,z^n = 
-\int^1_0 ds \frac{z}{1-zp} \log (1-zs)  \nonumber \\
\z\z 3\sum_{n=1}^{\infty}{2n\choose n}^{\!\!\!-1}
 \frac{1}{n}S_1 W_1\,z^n = 
-\int^1_0 ds \frac{z}{1-zp} \log (1-zs) \log \biggl(\frac{s}{(1-s)(1-zp)} 
\biggr)  \nonumber \\
\z\z \sum_{n=1}^{\infty}{2n\choose n}^{\!\!\!-1}
 \frac{1}{n} \biggl[3W_{1,1} + W_2 \biggr]
\,z^n = 
-\int^1_0 ds \frac{z}{1-zp} \biggl[ 2\Li_2(pz) + \nonumber \\
\z\z~~~\log (1-zs) \log \biggl(\frac{s}{1-s} \biggr) \biggr]  \nonumber \\
\z\z 3\sum_{n=1}^{\infty}{2n\choose n}^{\!\!\!-1}
 \frac{1}{n^2} W_1 \,z^n = 
-\int^1_0  \frac{ds}{p} \biggl[ \Li_2(s) - \Li_2(\frac{s}{1-zp})
+ \log (1-s)\log (1-zp) \nonumber \\
\z\z~~~- \frac{1}{2}\log^2 (1-zp) \biggr]  \nonumber \\
\z\z 3\sum_{n=1}^{\infty}{2n\choose n}^{\!\!\!-1}
 \frac{1}{n^2} S_1 W_1 \,z^n = 
\int^1_0  \frac{ds}{p} \biggl[ \Li_3(s) - \Li_3(\frac{s}{1-zp}) -
\log \biggl(\frac{s}{1-s}\biggr)\Li_2(s) \nonumber \\
\z\z ~~~
+ \log \biggl(\frac{s}{(1-s)(1-zp)} 
\biggr) \Li_2(\frac{s}{1-zp})
- \frac{1}{3}\log^3 (1-zp) + \frac{1}{2} \log (1-s)\log^2 (1-zp)
\nonumber \\
\z\z ~~~ -
\log (1-zp)\log (1-s)\log \biggl(\frac{s}{1-s} \biggr)
 \biggr]  \nonumber \\
\z\z \sum_{n=1}^{\infty}{2n\choose n}^{\!\!\!-1}
 \frac{1}{n^2} \biggl[3W_{1,1} + W_2 \biggr]
\,z^n = 
\int^1_0  \frac{ds}{p} \biggl[ 2\log (1-zp) \Li_2(pz) +4S_{1,2}(pz)
\nonumber \\
\z\z ~~~
-\log \biggl(\frac{s}{1-s}\biggr)
\biggl( \Li_2(s) - \Li_2(\frac{s}{1-zp}) \biggr)
+ \log (1-s)\log (1-zp) \biggr) \biggr] \nonumber \\
\z\z \sum_{n=1}^{\infty}{2n\choose n}^{\!\!\!-1}
 \frac{1}{n} S_1 \biggl[3W_{1,1} + W_2 \biggr]
\,z^n = 
\int^1_0 ds \frac{z}{1-zp} \Phi_1(z,p) \nonumber \\
\z\z 3\sum_{n=1}^{\infty}{2n\choose n}^{\!\!\!-1}
 \frac{1}{n}S^2_1 W_1\,z^n = 
\int^1_0 ds \frac{z}{1-zp} \biggl[ \Phi_2(z,p) + \Phi_1(z,p) 
\biggr] \nonumber \\
\z\z 3\sum_{n=1}^{\infty}{2n\choose n}^{\!\!\!-1}
 \frac{1}{n}S_2 W_1\,z^n = 
\int^1_0 ds \frac{z}{1-zp} \biggl[ \Phi_2(z,p) - \Phi_1(z,p) 
- \Phi_3(z,p) \biggr] \nonumber \\
\z\z \sum_{n=1}^{\infty}{2n\choose n}^{\!\!\!-1}
 \frac{1}{n} \biggl[W_3 - 3W_{1,2} \biggr] \,z^n = 
\int^1_0 ds \frac{z}{1-zp} \biggl[ \Phi_4 - \Phi_5 -\Phi_2 + \Phi_1
+\Phi_3  \biggr] \nonumber \\
\z\z \sum_{n=1}^{\infty}{2n\choose n}^{\!\!\!-1}
 \frac{1}{n} \biggl[3W_{1,(1+1)} + 2W_{2,1}\biggr] \,z^n = 
\int^1_0 ds \frac{z}{1-zp} \biggl[ \Phi_5(z,p) + \Phi_1(z,p) 
\biggr], \nonumber 
\end{eqnarray}
where 
\begin{eqnarray}
\z\z \Phi_1(z,p)= 
\log \biggl(\frac{s}{1-s}\biggr)
\biggr[ \Li_2(\frac{zs^2}{1-zp}) +\frac12\log^2 (\frac{1-zs}{1-zp}) +
\log (1-zp) \log (\frac{1-zs}{1-zp}) \biggr] \nonumber \\
\z\z ~~~ -2\biggr[ \Li_3(pz) -S_{1,2}(pz) -\log (1-zp)\Li_2(pz) \biggr] 
\nonumber \\ 
\z\z \Phi_2(z,p)= 
\Li_3(\frac{zs^2}{1-zp}) +S_{1,2}(\frac{zs^2}{1-zp})
-S_{1,2}(pz) -\log (1-zp) \biggl[ \Li_2(\frac{zs^2}{1-zp}) + \Li_2(pz)
\biggr] 
\nonumber \\    \z\z~~~
- \frac{1}{2} \log^3 (1-zp) \nonumber \\
\z\z \Phi_3(z,p)= 2S_{1,2}(\frac{zs^2}{1-zp})
-2\Li_3(zp) +2 \log (\frac{1-zs}{1-zp})\Li_2(\frac{zs^2}{1-zp}) 
 + \frac{1}{3} \log^3 (1-zs) \nonumber \\
\z\z~~~-  \log^2 (1-zs) \log (1-zp) + 
\log (1-zs) \log^2 (1-zp)  \nonumber \\
\z\z \Phi_4(z,p)= 2\Li_3(\frac{zs^2}{1-zp})
- \Li_2(\frac{zs^2}{1-zp})\log \biggl(\frac{s}{(1-s)(1-zp)}\biggr)
+2\Li_3(pz) + S_{1,2}(pz)
    \nonumber \\
\z\z ~~~
+ \frac{5}{6} \log^3 (1-zp) \nonumber \\
\z\z 9\Phi_5(z,p)= 2\biggl[ \Li_3(pz) - 6S_{1,2}(pz)
-3\log (1-zp)\Li_2(pz) - \frac{1}{6} \log^3 (1-zp) \biggr]
  \nonumber \\
\z\z ~~~
- 4\biggl[ \Li_3(\frac{zs^2}{1-zp}) +2S_{1,2}(\frac{zs^2}{1-zp}) \biggr]
- \Li_2(\frac{zs^2}{1-zp}) \log \biggl(\frac{s}{(1-s)(1-zp)^2}\biggr)\,. 
 \nonumber
\end{eqnarray}

%%%%%%%%%%%%%%%%
%%%%%%%%%%%%%%%%  Figure Captions 
%%%%%%%%%%%%%%%%
\newpage
\thispagestyle{empty}

\begin{center}
\Large\bf Figure captions
\end{center}

\noindent
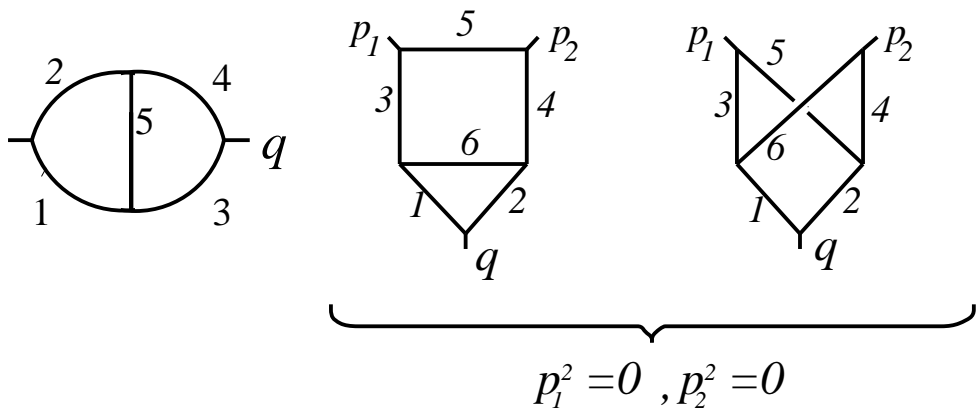
{\bf Fig. 1} : 2-loop selfenergy- and vertex diagrams considered
in the paper. Line numbering and kinematics.\\[3mm]

\noindent
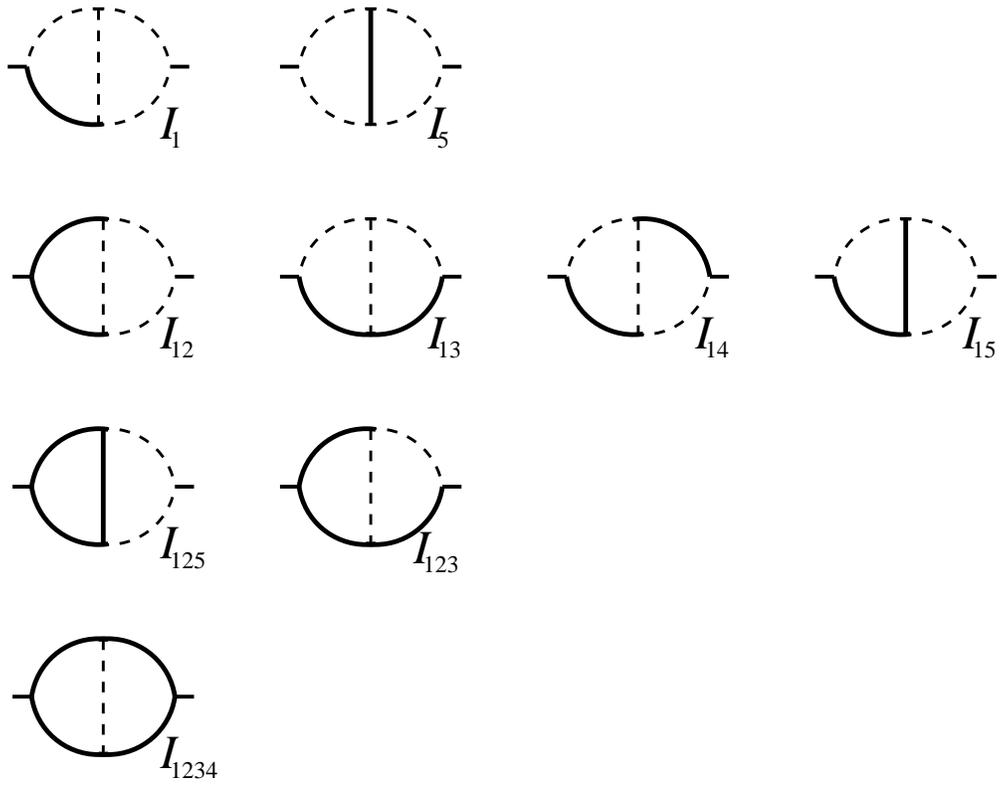
{\bf Fig. 2} : 2-loop selfenergy diagrams evaluated in this work.
Solid lines denote propagators with the mass $M$; dashed lines
denote massless propagators.\\[3mm]

\noindent
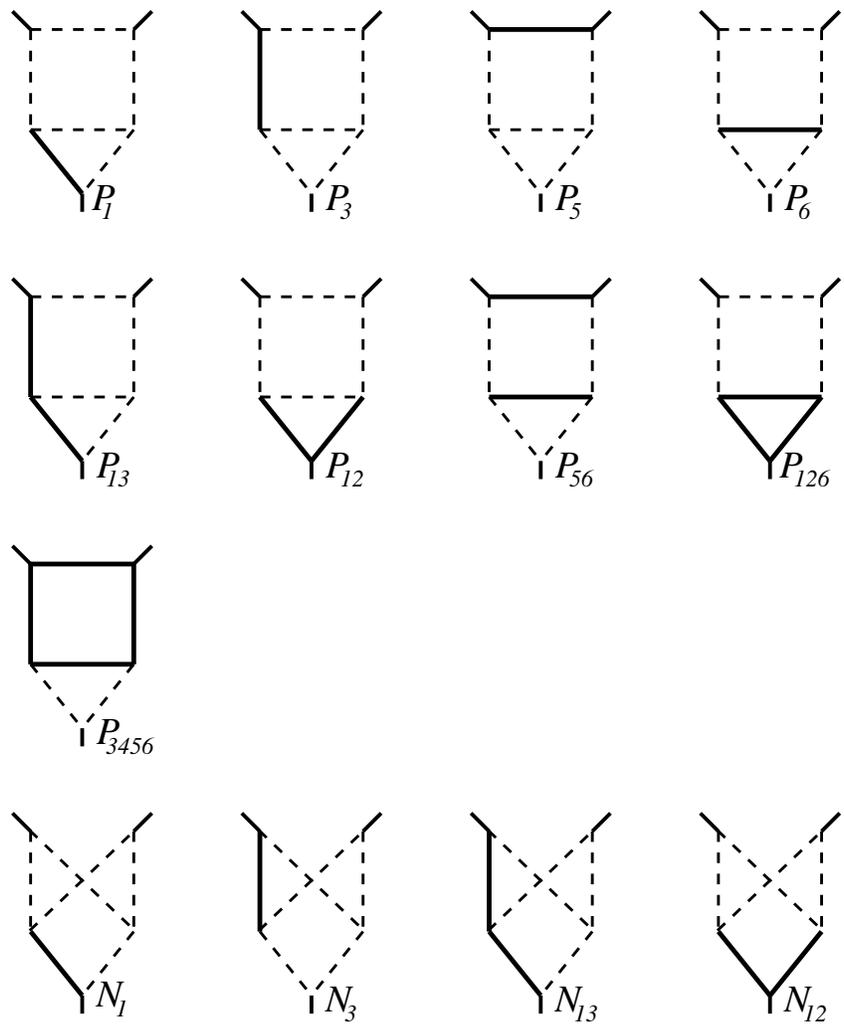
{\bf Fig. 3} : 2-loop vertex diagrams evaluated in this work.
Kinematics as in Fig. 1 ($p_1^2=p_2^2=0$).
Solid lines denote propagators with mass $M$; dashed lines
denote massless propagators.


\begin{thebibliography}{99}

\bibitem{Kaellen}
G. K\"all\'en and A. Sabry, Dan.Mat.Fys.Medd. 29 (1955) No.17;
R. Barbieri, J.A. Mignaco and E. Remiddi, Nuov.Cim. {\bf A11} (1972) 824, 865.

\bibitem{BroadhurstZP47}
D.J. Broadhurst, Z.Phys. {\bf C47} (1990) 115.

\bibitem{BFT}
D.J. Broadhurst, J. Fleischer and O.V. Tarasov, Z.Phys. C60 (1993) 287.

\bibitem{SE2}
S. Bauberger et al., Nucl.Phys. {\bf B434} (1995) 383;
V. Borodulin and G. Jikia, Phys.Lett. {\bf B391} (1997) 434. 

\bibitem{SE2bis}
R. Scharf and J.B. Tausk, Nucl.Phys. {\bf B412} (1994) 523.

\bibitem{Ghinculov}
A. Ghinculov and J.J. van der Bij, Nucl.Phys. {\bf B436} (1995) 30;
V. Borodulin, G. Jikia, Phys.Lett. {\bf B391} (1997) 434.

\bibitem{DEM}
A.V. Kotikov,  Phys. Lett. {\bf B254} (1991) 158;
{\bf B259} (1991) 314;
{\bf B267} (1991) 123 (Err. {\bf B295} (1992) 409);
In: {\it Artificial Intelligence and Expert Systems for
High Energy and Nuclear Physics}, (Oberamergau, Germany, 1993) p.453;
JINR preprint E2-93-414.

\bibitem{FKV}
J. Fleischer, A.V. Kotikov and O.L. Veretin, 
   Phys.Lett. {\bf B417} (1998) 163.

\bibitem{bubbles}
D.J. Broadhurst, hep-th/9803091. 

\bibitem{asymptotic}
F.V.~Tkachov, Preprint INR P-0332, Moscow (1983); P-0358, Moscow 1984;
K.G.~Chetyrkin,
Teor. Math. Phys. {\bf 75} (1988) 26; ibid {\bf 76} (1988) 207;
Preprint, MPI-PAE/PTh-13/91, Munich (1991);
V.A.~Smirnov, Comm. Math. Phys.{\bf 134} (1990) 109;
{\it Renormalization and Asymptotic Expansions},
Birkh\"auser, Basel, 1991.

\bibitem{ft}
J.~Fleischer and O.V.~Tarasov,
{  Z.Phys.}, {\bf C 64} (1994) 413;
J. Fleischer, V. A. Smirnov and O. V. Tarasov,
Z.Phys.{\bf C74} (1997) 379;
J.Fleischer {\it et al.}, Eur.Phys.J. C2 (1998) 747.

\bibitem{DT}
A.I. Davydychev and J.B. Tausk, Nucl.Phys. B397 (1993) 123. 

\bibitem{propagator}
A.I. Davydychev, V.A. Smirnov and J.B. Tausk,
Nucl. Phys.{\bf B410} (1993) 325;
F.A. Berends, A.I. Davydychev and  V.A. Smirnov,
Nucl. Phys.{\bf B478} (1996) 59;
L.V. Avdeev and M.Yu. Kalmykov,
Nucl. Phys.{\bf B502} (1997) 419;
G. Weiglein, hep-ph/9711254.

\bibitem{propagatorbis}
F.A. Berends, A.I. Davydychev, V.A. Smirnov and J.B. Tausk,
Nucl. Phys.{\bf B439} (1995) 536;

\bibitem{Yndurain}
F.J. Yndurain, {\it Quantum Chromodynamics}
  (Springer-Verlag New York Inc., 1983).

\bibitem{polylogarithms}
L. Lewin, {\it Polylogarithm and associated functions}
  (North-Holland, Amsterdam, 1981);\\
N. Nielsen, Nova Acta Leopold. 90 (1909) 123.

\bibitem{VermRh}
J.A.M. Vermaseren, NIKHEF-98-14, hep-ph/9806280.

\bibitem{Bauberger}
S. Bauberger and M. B\"ohm, Nucl.Phys. {\bf B445} (1995) 25.

\bibitem{DIS}
D.I. Kazakov and  A.V. Kotikov, Nucl.Phys. {\bf B307} (1988) 721;
Theor.Math.Phys. {\bf 73} (1988) 1264; A.V. Kotikov, 
Theor.Math.Phys. {\bf 78} (1989) 134. 

\bibitem{FL}
D.I. Kazakov and  A.V. Kotikov, Phys.Lett. {\bf B291} (1992) 171.

\bibitem{Remiddi}
E. Remiddi, Nuovo Cim. {\bf 110A} (1997) 1435.

\bibitem{bubbl}
A.V. Kotikov, hep-ph/9807440.

\bibitem{FORM}
J.A.M. Vermaseren, {\it Symbolic Manipulation with FORM}
(CAN, Amsterdam, The Netherlads, 1991).

\bibitem{Devoto}
A.~Devoto and D.~W.~ Duke, La Rivista del Nuovo Cimento, 
Vol. 7, No. 6, pg.1 (1984).

\bibitem{bypart}
K.G. Chetyrkin and F.V. Tkachov, Nucl. Phys. {\bf B192} (1981) 159;
F.V. Tkachov,  Phys. Lett. {\bf B100} (1981) 65; 
A.N. Vassiliev, Y.M. Pis'mak and Y.P. Honkonen, Theor. Math. Phys. 
{\bf 47} (1981) 465.

\bibitem{Smirnov}
V.A.~Smirnov, Phys. Lett. {\bf B404} (1997) 101;

\end{thebibliography}
\end{document}